\documentclass{article}

\newif\ifpdf
\ifx\pdfoutput\undefined
\pdffalse 
\else
\pdfoutput=1 
\pdftrue
\fi

\ifpdf
\usepackage[pdftex]{graphicx}
\else
\usepackage{graphicx}
\fi

\ifpdf
\DeclareGraphicsExtensions{.pdf, .jpg}
\else
\DeclareGraphicsExtensions{.eps, .jpg}
\fi

\usepackage{graphics}
\usepackage{graphicx}
\usepackage[centertags]{amsmath}
\usepackage{amsfonts}
\usepackage{amsthm}
\usepackage{amssymb}
\usepackage{url}

\begin{document}

\title{Determining a regular language by glider-based structures called {\it phases f$_i$\_1} in Rule 110}
\author{Genaro Ju\'arez Mart\'{\i}nez$^1$, Harold V. McIntosh$^2$, \\
Juan C. Seck Tuoh Mora$^3$ and Sergio V. Chapa Vergara$^4$}
\date{November 1, 2006\footnote{by publish in special issue of the {\it Jornal of Cellular Automata}}}

\maketitle

\begin{centering}

$^1$ Faculty of Computing, Engineering and Mathematical Sciences, University of the West of England, Bristol, United Kingdom. \\ \url{genarojm@correo.unam.mx} \\

$^2$ Departamento de Aplicaci\'on de Microcomputadoras, Instituto de Ciencias, Universidad Aut\'onoma de Puebla, Puebla, M\'exico. \\ \url{mcintosh@servidor.unam.mx} \\

$^3$ Centro de Investigaci\'on Avanzada en Ingenier\'{\i}a Industrial, Universidad Aut\'onoma del Estado de Hidalgo Pachuca, Hidalgo, M\'exico. \\ \url{jseck@uaeh.edu.mx} \\

$^4$ Departamento de Computaci\'on, Centro de Investigaci\'on y de Estudios Avanzados del Instituto Polit\'ecnico Nacional, M\'exico. \\ \url{schapa@cs.cinvestav.mx} \\

\end{centering}

\maketitle

\begin{abstract}

Rule 110 is a complex elementary cellular automaton able of supporting universal computation and complicated collision-based reactions between gliders. We propose a representation for coding initial conditions by means of a finite subset of regular expressions. The sequences are extracted both from de Bruijn diagrams and tiles specifying a set of phases f$_i$\_1 for each glider in Rule 110. The subset of regular expressions is explained in detail.

\end{abstract}

\section{Introduction}

The study of the binary-state one-dimensional cellular automaton Rule 110 has had a certain attention before and after the demonstration that its evolution space can bear universal computable processes (see \cite{kn:Cook04,kn:Wolf02}).

Another important and complementary result to the previous one was obtained by Turlough Neary and Damien Woods showing that the problem of predicting $t$ steps in Rule 110 is P-complete. Some interesting reductions of Turing machines are displayed in \cite{kn:NW06}. On the other hand, Kenichi Morita have finish a complicate and new results in cyclic tag systems \cite{kn:Kenichi07,kn:Kenichi}. Mainly over the ``halt'' problem in this systems.

The diversity of problems in Rule 110 and its possible applications in different fields determine the necessity of formalizing a representation for coding systematically the evolution rule; for easily constructing initial conditions which define a control of the gliders (particles or mobile localizations) taking part in simple or complicated complex operations.

In the present paper we report a set of sequences based on gliders that can be represented as regular expressions codified in initial conditions offering a way to manipulate the glider system in Rule 110.

The paper gives a brief introduction to Rule 110 and its glider system. Later it presents a small review on regular languages, de Bruijn diagrams and tiles. Finally it explains how the expressions are calculated for all the gliders up to now known in Rule 110 (without extensions), illustrating a simple procedure to handle collisions between gliders and depicting some relevant constructions.

A pertinent mention is that this set of regular expressions has been successfully applied in some of our previous results in Rule 110 \cite{kn:JM01,kn:JMS03,kn:JMS06,kn:JMSa,kn:JSM}.

\section{Basic notation}

Rule 110 is a cellular automaton of order $(k=2,r=1)$ (Wolfram's notation) evolving in one dimension, where $k$ determines the number of states of an alphabet $\Sigma$ and $r$ is the number of cells considered both to the left and to the right side with regard of a central cell.

Particularly, Rule 110 can produce a wide variety of gliders on a periodic background called ``ether'' by Matthew Cook \cite{kn:Cook04,kn:Cook99}. Thus, Rule 110 belongs to Class IV in Wolfram's classification.

The local function determining the behavior of Rule 110 is:

\begin{table}[th]
\centering
\begin{tabular}{cc}
$\varphi(0,0,0) \rightarrow 0$ & $\varphi(1,0,0) \rightarrow 0$ \\
$\varphi(0,0,1) \rightarrow 1$ & $\varphi(1,0,1) \rightarrow 1$ \\
$\varphi(0,1,0) \rightarrow 1$ & $\varphi(1,1,0) \rightarrow 1$ \\
$\varphi(0,1,1) \rightarrow 1$ & $\varphi(1,1,1) \rightarrow 0$ 
\end{tabular}
\caption{Evolution rule 110.}
\label{regla-110}
\end{table}

The evolution rule is expressed in binary notation 01101110 (representing the decimal number 110). The global evolution of the automaton is defined starting from linear array of cells each containing one state of $\Sigma$; taking every cell $x_i$ as a central one, we evaluate the value of its corresponding neighborhood to determine the new central element in the following generation:

\begin{center}
$\varphi(x_{i-1}^{t},x_{i}^{t},x_{i+1}^{t}) \rightarrow x_{i}^{t+1}$.
\end{center}

Time $t$ is discrete and there is a simultaneous evaluation of each $x_i$ in the array, i. e., parallel mappings generate the following array, determining the evolution space $\Sigma^{\mathcal Z}$.

\begin{figure}[th]
\centerline{\includegraphics[width=4.8in]{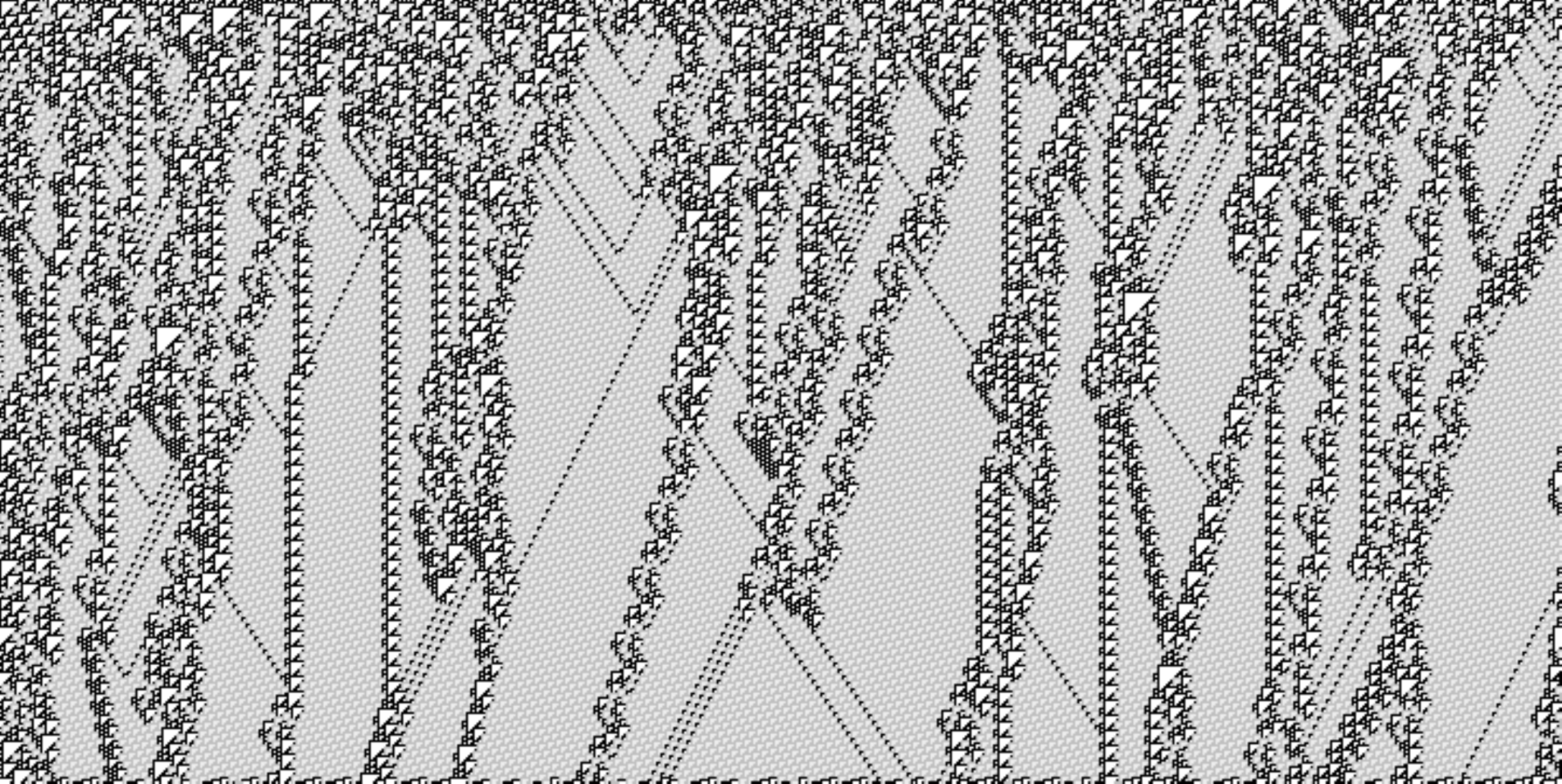}}
\caption{Random evolution in Rule 110.}
\label{evol_aleatoria}
\end{figure}

Figure~\ref{evol_aleatoria} shows a typical random evolution in Rule 110 with an initial density of 0.5 in an array with 723 cells for 363 steps. In the evolution we have applied a filter identifying the ether configurations allowing a clear recognizing of the gliders present in this example.

Once established the existence of gliders (particles or mobile localizations) we must classify them and determine their properties.

\section{Glider system in Rule 110}

\begin{figure}[th]
\centerline{\includegraphics[width=6in]{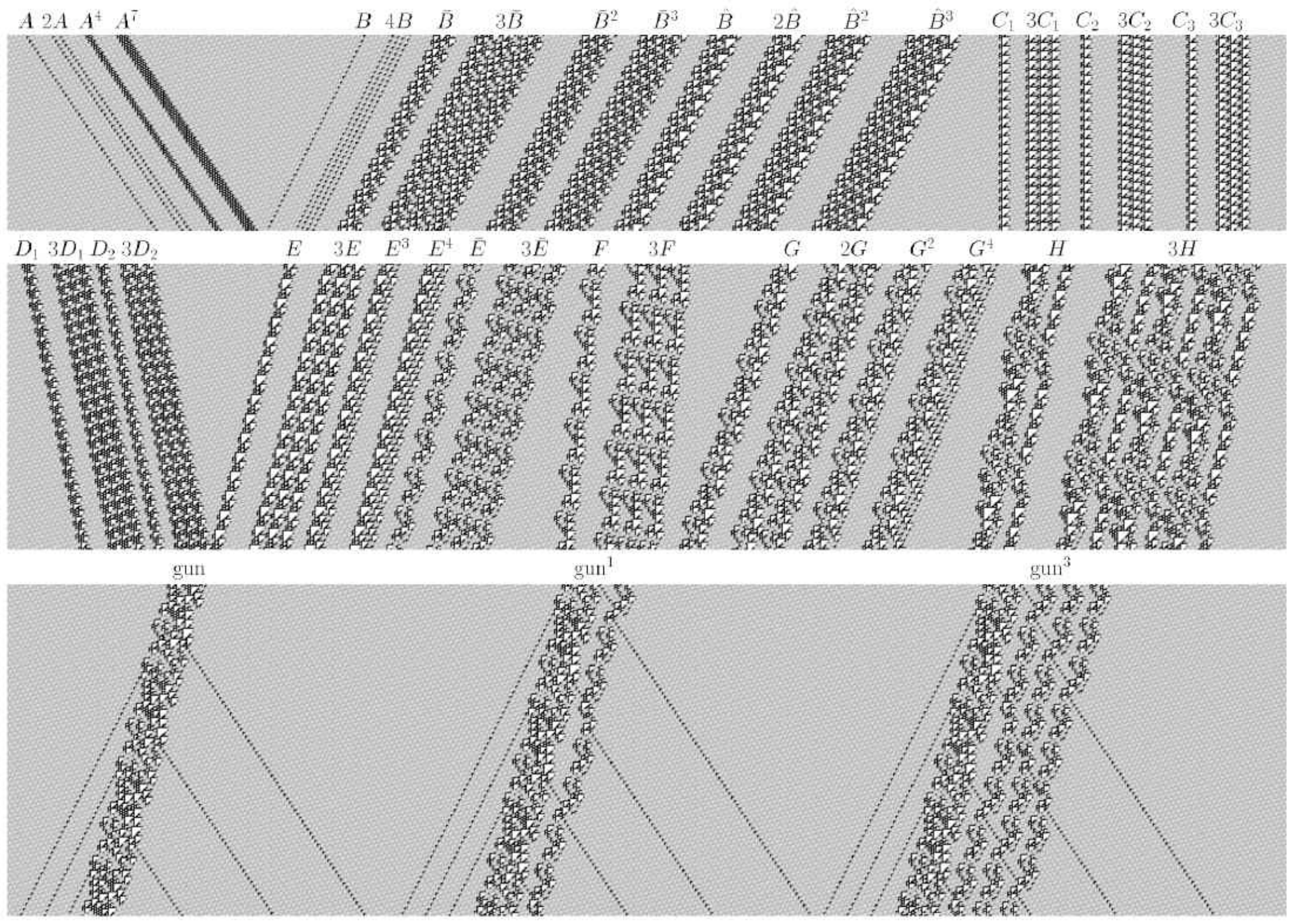}}
\caption{Glider classification in Rule 110.}
\label{listaCookgliders}
\end{figure}

In this section we show all the gliders until now known in Rule 110. Let us use the classification proposed by Cook \cite{kn:Cook04} from now on (illustrated in Figure~\ref{listaCookgliders}).

Gliders are presented in two forms: as a simple structure and as extensions or packages of them (one example showed in Figure~\ref{listaCookgliders}). Each glider with superscript $n \in \mathbb Z^+$ represents that it can arbitrarily extend; extensions to the left are defined in $\bar{B}$ and $\hat{B}$ gliders, and extensions to the right are with $E$ and $G$ gliders. At the end of the list there is an extended glider gun, where the extension is originated by $\bar{E}$ gliders. Also, we can see examples of extended gliders and packages with their respective notation.

In the evolution space of Rule 110 we can see three trajectories for the gliders. A shift from left to right is made by $A$, $D_1$ and $D_2$ gliders and a shift from right to left is realized by $B$, $\bar{B}$, $\hat{B}$, $E$, $\bar{E}$, $F$, $G$, $H$ gliders and the glider gun. The last trajectory is with gliders which does not have a shift, $C_1$, $C_2$ and $C_3$ gliders. Each glider has a period determined by the number of generations among shifts letting the same sequence or the change from $x_{i}$ to $x_{i+d}$ or $x_{i-d}$, where $d \in \mathbb Z$ represents the number of places crossed in every period.

An important indication is that the set of regular expressions $\Psi_{R110}$ describing all the gliders in Rule 110 does not include extensions or packages of them, it is only for simple gliders. On trying to enumerate all those extensions or packages, the set of expressions grows in different modules; therefore, the number of sequences $w$ in the set is the union of the periods for every glider:

\begin{equation}
\Psi_{R110} = \bigcup_{i=1}^{p}{w_{i,g}} \mbox{ } \forall \mbox{ } (w_i \in \Sigma^* \wedge g \in \cal G)
\end{equation}

\noindent where $\cal G$ is the whole set of gliders in Rule 110 and $p \geq 3$ is the corresponding period. This way, we can speak of a regular language $L_{R110}$ that is constructed from the expressions of $\Psi_{R110}$. We must notice that this language is a subset of the whole language in Rule 110, that is, it is only the one defined by the expressions representing gliders, then we have:

\begin{equation}
L_{R110} = \{w | w \in \Psi_{R110} \mbox{ operating under the basic rules: } \cdot,+,*\}.
\end{equation}

Language $L_{R110}$ is based on the regular expressions $\Psi_{R110}$ determining each glider; a remarkable comment is that $L_{R110}$ has not been published or explained by other authors.

$L_{R110}$ is established by the de Bruijn diagrams and the characterization of the tiles, where both have been analyzed for defining useful features called ``phases.'' The phases indicate with precision both the position and the exact moment where each glider must be positioned into a given initial condition.

When applying the set of regular expressions and their basic operations we are able to construct desired initial conditions which yield evolutions with important characteristics; the main interest is to control and produce collisions among gliders. In this way $L_{R110}$ is a powerful tool to codify initial conditions in Rule 110, and this subset has been implemented in a computer system. Immediate applications with relevant results in the study of Rule 110 has been performed over hundreds, thousands, millions and thousands of million of cells, as we shall see in the following section.

Now we describe the properties of each glider in different aspects such as: name, periodic margins, speed, width of the structure and the cap by glider in the evolution space (see Table~\ref{margenesgliders}).

\begin{table}[th]
\centering
\small
\begin{tabular}{||c|c|c|c|c|c|c|c||}
\hline \hline
                 & \multicolumn{4}{c|}{margins} & & &  \\
structure & \multicolumn{4}{c|}{left - right} & $v_{g}$ & width & cap \\
\cline{2-5}
 & $ems$ & $oms$ & $ems$ & $oms$ & & & \\
\hline \hline
$e_{r}$ & . & 1 & . & 1 & 2/3 $\approx$ 0.666666 & 14 & T \\
\hline
$e_{l}$ & 1 & . & 1 & . & -1/2 = -0.5 & 14 & T \\
\hline
$A$ & . & 1 & . & 1 & 2/3 $\approx$ 0.666666 & 6 & T \\
\hline
$B$ & 1 & . & 1 & . & -2/4 = -0.5 & 8 & P \\
\hline
$\bar{B}^{n}$ & 3 & . & 3 & . & -6/12 = -0.5 & 22 & T \\
\hline
$\hat{B}^{n}$ & 3 & . & 3 & . & -6/12 = -0.5 & 39 & T \\
\hline
$C_{1}$ & 1 & 1 & 1 & 1 & 0/7 = 0 & 9-23 & P \\
\hline
$C_{2}$ & 1 & 1 & 1 & 1 & 0/7 = 0 & 17 & P \\
\hline
$C_{3}$ & 1 & 1 & 1 & 1 & 0/7 = 0 & 11 & P \\
\hline
$D_{1}$ & 1 & 2 & 1 & 2 & 2/10 = 0.2 & 11-25 & P \\
\hline
$D_{2}$ & 1 & 2 & 1 & 2 & 2/10 = 0.2 & 19 & P \\
\hline
$E^{n}$ & 3 & 1 & 3 & 1 & -4/15 $\approx$ -0.266666 & 19 & P \\
\hline
$\bar{E}$ & 6 & 2 & 6 & 2 & -8/30 $\approx$ -0.266666 & 21 & P \\
\hline
$F$ & 6 & 4 & 6 & 4 & -4/36  $\approx$ -0.111111 & 15-29 & P \\
\hline
$G^{n}$ & 9 & 2 & 9 & 2 & -14/42 $\approx$ -0.333333 & 24-38 & P \\
\hline
$H$ & 17 & 8 & 17 & 8 & -18/92 $\approx$ -0.195652 & 39-53 & P \\
\hline
glider gun & 15 & 5 & 15 & 5 & -20/77 $\approx$ -0.259740 & 27-55 & P \\
\hline \hline
\end{tabular}
\caption{Properties to each glider in Rule 110.}
\label{margenesgliders}
\end{table}

In Table~\ref{margenesgliders}, column {\it structure} represents the name of the glider or periodic structure. The following four columns labeled {\it margins}, indicate the number of periodic margins in each glider. The margins are divided in margins with even values `{\it ems}` and odd values `{\it oms}'  which are distributed as well in two groups: left and right, because gliders has even and odd margins in their left or right borders (or superior and inferior ones). Particularly, the properties of the margins are explained in subsection 4.4, discussing their origins, interpretations and representations.

Column $v_g$ $\forall$ $g \in \cal G$ indicates the speed of each glider, where it is calculated dividing the shift $d$ between its period $p$. The three types of trajectories are identified in this column. Positive speed indicates a shift to the right, negative speed a shift to the left and a zero speed tells that the glider does not have a shift.

Column $width$ indicates the minimum and maximum number of necessary cells for determining a periodic chain in the linear array forming a glider or another periodic structure. For example, for the $C_1$ glider we have two values, this means that with nine or twenty-three cells may define this glider in the initial condition.

The last column {\it cap} indicates the gliders able to completely cover the evolution space of Rule 110. The cap can be total `T' or partial `P,' where total cap implies a glider which does not need additional tiles to completely cover the evolution space. A partial cap describes that at least the intervention of another tile is necessary so that the glider and the new tile can completely cover the evolution space. This representation is oriented to the problem established by McIntosh to cover the space with different tiles and to find the combination of gliders fulfilling this condition.

Thus, another tendency in the research is represented by looking for possible complex constructions through tiles rather than using initial conditions. From an initial set of tiles $X$ we can construct a family of different sets $X_i$ so that, $\{ X \subset  \ldots  \subset X_i \subset X_{i+1} \subset \ldots \}$ and each set must produce a different pattern, hence we will make operations with the tiles in the cartesian plane as in a puzzle but without violating the valid connections determined by Rule 110. Therefore, in the sense of Hao Wang, we can find a composition of different sets $X_{i}$ to implement a sequence of tiles being operated by a logical function, describing another way of universal computation based on these constructions \cite{kn:GS82}.

\section{Determining a glider-based regular language in Rule 110}

This section explains the definition and representation of phases in the evolution space of Rule 110. The analysis starts with the description of the evolution space by tiles evolutions and we applied de Bruijn diagrams \cite{kn:Mc91,kn:Voor96} to specify the finite subset of glider-based regular expressions. Both approaches give origin to the interpretation of ``phases'' in Rule 110; once determined the phases, a procedure is explained to control specific collisions among gliders codified into initial conditions applying the subset $\Psi_{R110}$ of regular expressions establishing a regular language $L_{R110}$.

\subsection{Tiles in Rule 110}

A {\it plane of tiles} $\mathcal T$ is a countable family of closed sets $\mathcal T = \{T_{0},T_{1},\ldots\}$ covering the plane without intervals or intersections \cite{kn:GS82}. Defined as a join of sets (called a mosaic $\mathcal T$):

\begin{equation}
\mathcal T = \bigcup_{i=0}^{n}{T_{i}} \mbox{ $\forall$ } n \in \mathbb Z^+_0
\end{equation}

\noindent The ``plane'' is the Euclidian plane $\mathbb Z \times \mathbb Z$ in elementary geometry.

\begin{figure}[th]
\centerline{\includegraphics[width=4.8in]{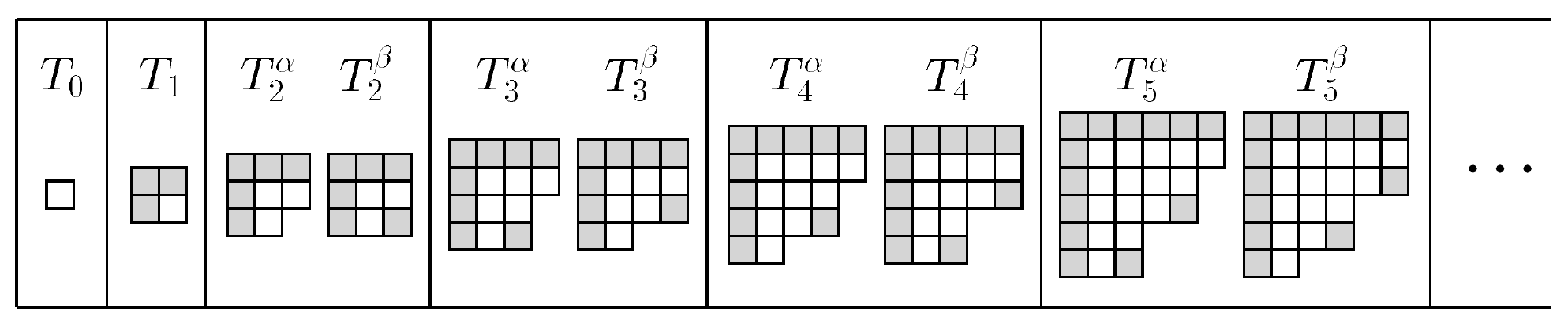}}
\caption{Two types of tiles in Rule 110: $\alpha$ and $\beta$.}
\label{tiles}
\end{figure}

Rule 110 covers the evolution space through different sets of triangles $T_n$ $\forall$ $n \in \mathbb Z^+_0$, where $n$ represent the size of the triangle counting the cells in some of its internal sides. The tiles are divided in two sets: $\alpha$ and $\beta$ $\forall$ $n \geq 2$ \cite{kn:Mc99} (each set $\alpha$ or $\beta$ determines its own countable family of tiles where $|\{T_{n}^{\alpha}\}| = |\{T_{n}^{\beta}\}|$, as illustrates Figure~\ref{tiles}). For example, different $\alpha$ and $\beta$ tiles are present in the construction of both the $H$ glider and the glider gun (see Figure 4 in \cite{kn:JMS06}).

We can represent a $T_{0}$ tile by state 0; with this when the initial configuration is covered by the expressions: 0*, 1* and (10)*, the evolution space is established by a homogenous evolution with state 0 (or tile $T_{0}$). Nevertheless, the behavior is not the same for tiles $T_{1}, T_{2}^{\alpha}, T_{2}^{\beta}, T_{3}^{\alpha}, T_{3}^{\beta}, \ldots, T_{n}^{\alpha}, T_{n}^{\beta}, \ldots$. The evolution space can be covered by any $T_{n}$ tiles for $0 \leq n \leq 4$. Thus for $n \geq 5$ the evolution space is covered by at least two $T_{n}$ tiles. Let $T_{i}$ and $T_{j}$ $\in \mathcal T$ where $i \neq j$, then both sets cannot operate in the plane under the function of Rule 110 if they cover the space partially (gaps) or overlap in their cells.

Another question is to know the largest tile that Rule 110 can construct in its evolution space. At the present time, the limit is established by a $T_{45}$ tile \cite{kn:Mc00}; therefore, Rule 110 cannot construct a greater mosaic. At the moment there is a way to produce $T_{n}$ tiles where $0 \leq n \leq 33$; tiles $T_{43}$, $T_{44}$ and $T_{45}$ were calculated through a specialized search determining the ancestors for each tile \cite{kn:Mc00}. Finally, other open problem is to determine a construction for tiles in the interval $34 \leq n \leq 42$. 

Thus, the tile family $\{T_{n}^{\alpha}\}$ and $\{T_{n}^{\beta}\}$ allows a detailed description of the evolution space in Rule 110 through their sets: $\alpha$ and $\beta$. A second important point is that the tiles establish properties by the periodic margins in their recurrent structures (gliders and ether). Their interpretation is very important to derive the phases, including non-periodic structures.

As we said before, if from an initial set $X$ a family of different sets $X_i$ is defined so that $\{ X \subset  \ldots  \subset X_i \subset X_{i+1} \subset \ldots \}$, a function $\Gamma: \cal T^* \rightarrow \cal T^*$ can be defined. Thus, we have each $\bigcup_{i=0}^{n}{T_{i}} = X_i \subset \mathcal G$ where $X_i = \mathcal G_N$ $\vee$ $\mathcal G_C$.\footnote{Where $\mathcal G_N$ represent natural gliders and $\mathcal G_C$ represent compound gliders \cite{kn:JMS06}.}

\subsection{Regular expressions}

Several interesting problems rise in the study of formal languages; one of them is to determine the type of language derived and to which class belongs. This hierarchy is well-known and established by Chomsky's classification. We shall study languages determined by regular sets, since the set of expressions determined by each glider in Rule 110 can be associated to a particular regular expression. Thus, some concepts of finite state machines are needed.

The finite automaton is a mathematical model with a system of discrete inputs and outputs;  the system can be placed  in one of a finite set of states. This state has the information of the received inputs necessary to determine the behavior of the system with regard of subsequent inputs. Formally, a finite automaton $M$ consists of a finite set of states and a set of transitions among states induced by the symbols selected from some alphabet. For each symbol there is a transition form one state to other (it can return to the same one); there is an initial state where the automaton stars and some states are designated as final ones or acceptance states  \cite{kn:HU79}.

A directed graph called a {\it transition diagram} is associated with a finite automaton as follows: the vertices of the graph correspond to the states of the automaton; for a transition from state $i$ to state $j$ produced by an input symbol, there is an edge labeled by this symbol from $i$ to $j$ in the transition diagram. The finite automaton accepts a chain $w$ if the analogous transition sequence leads from the initial state to a final one (or acceptation).

A {\it language accepted} by $M$, represented by $L(M)$, it is the set $\{w | w$ is accepted by $M\}$. The type of languages accepted by a finite automaton is important because they complement the analysis established with regular expressions. Historically an important relation was established by S. C. Kleene demonstrating that regular expressions can be expressed by a finite automaton and vice versa, i. e., they are equivalent representations \cite{kn:Mins67}. In other words, a language is a {\it regular set} if it is accepted by some finite automaton. The accepted languages by finite automata are described by expressions known as {\it regular expressions}; particularly, the accepted languages by finite automata are indeed the class of languages described by regular expressions.

The sets of {\it regular expressions} on an alphabet are defined recursively as \cite{kn:HU79}:

\begin{enumerate}
\item $\phi$ is the regular expression representing the empty set.
\item $\epsilon$ is the regular expression describing the set $\{\epsilon\}$.
\item For each symbol $a \in \Sigma$, $a$ is a regular expression depicting the set $\{a\}$.
\item If $a$ and $b$ are regular expressions representing languages $A$ and $B$ respectively, then $(a+b)$, $(ab)$, and $(a^*)$ are regular expressions representing $A \cup B$, $AB$ and $A^*$ respectively.
\end{enumerate}

When it is necessary to distinguish between a regular expression $a$ and the language determined by $a$, we shall use $L_a$.

The formal languages theory provides a way to study sets of chains from a finite alphabet. The languages can be seen as inputs of some classes of machines or like the final result from a typesetter substitution system i.e., a generative grammar into the Chomsky's classification \cite{kn:Hurd87}.

\begin{table}[th]
\centering
\begin{tabular}{|c|c|c|}
\hline
language & structure \\
\hline \hline
recursively enumerated & Turing machine \\
\hline
context sensitive & linear bounded automata \\
\hline
context free & pushdown automata \\
\hline
regular & finite automata \\
\hline
\end{tabular}
\caption{Language classes.}
\label{herarquia-Chomsky}
\end{table}

The basic model necessary for the languages of these machines (and for all computation), is the Turing machine; the machines recognizing each family of languages are described as a Turing machine with restrictions. The relevance of associating a machine or system to resolve each type of language is for establishing a classification (Table~\ref{herarquia-Chomsky} of \cite{kn:Hurd87}).

Some languages are established by regular sets; although we can take all the words recognized by the de Bruijn diagram, we just need those chains representing a structure in Rule 110, to manipulate the evolution space with constructions of particles or gliders. Regular sets can be recognized by machines with finite memory (finite state machines) and may be generated by linear right (or left) grammars. Another way to represent chains in a regular language is by regular expressions.\footnote{Examples and properties of the formal languages, grammars, finite state machines, Turing machines and equivalent systems can be consulted in \cite{kn:Arb69,kn:HU79,kn:Mins67,kn:Dav82,kn:Ston73,kn:Tur36}.}

The regular language $L_{R110}$ is restricted to gliders in Rule 110. The application of this regular subset allows to solve some important problems, on defining initial conditions codified by phases; offering as well a powerful tool to codify the evolution space of Rule 110.\footnote{The regular language $L_{R110}$ does not imply that the evolution of Rule 110 is regular in the sense of limit sets \cite{kn:Wolf84,kn:Hurd87,kn:Nord89}, because $L_{R110}$ is only conserved in the composition of the initial conditions.}

\subsection{De Bruijn diagrams}

De Bruijn diagrams \cite{kn:Mc91,kn:Voor96,kn:Mc07} are very adequate for describing evolution rules in one-dimensional cellular automata, although originally they were used in shift-register theory (the treatment of sequences where their elements overlap each other). We shall explain de Bruijn diagrams illustrating their constructions for determining chains $w$ defining a pair of gliders in $\mathcal G$, the set of gliders in Rule 110.

For an one-dimensional cellular automaton of order $(k,r)$, the de Bruijn diagram is defined as a directed graph with $k^{2r}$ vertices and $k^{2r+1}$ edges. The vertices are labeled with the elements of the alphabet of length $2r$. An edge is directed from vertex $i$ to vertex $j$, if and only if, the $2r-1$ final symbols of $i$ are the same that the $2r-1$ initial ones in $j$ forming a neighborhood of $2r+1$ states represented by $i \diamond j$. In this case, the edge connecting $i$ to $j$ is labeled with $\varphi(i \diamond j)$ (the value of the neighborhood defined by the local function) \cite{kn:Voor96,kn:Voor06}.

\begin{figure}[th]
\centerline{\includegraphics[width=1.6in]{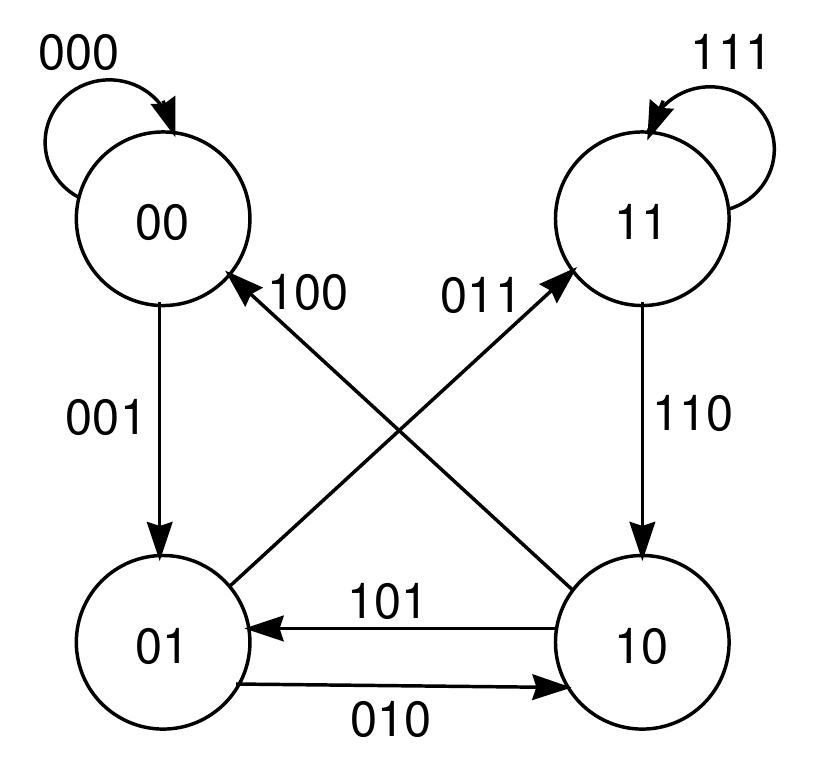}}
\caption{Generic de Bruijn diagram for a cellular automaton (2,1).}
\label{deBruijnGenerico}
\end{figure}

The connection matrix $M$ corresponding with  the de Bruijn diagram is as follows:

\begin{equation}
	M_{i,j} = \left\{\begin{array}{ll}
			        1& \mbox{if } j = ki, ki+1, \hdots, ki+k-1 \mbox{ (mod } k^{2r}) \\
		           	 0 & \mbox{in other case} \\
		       \end{array}
			\right.
\label{eq-Bruijn}
\end{equation}

Module $k^{2r}=2^{2}=4$ represent the number of vertices in the de Bruijn diagram and $j$ must take values from $k*i=2i$ to $(k*i)+k-1=(2*i)+2-1=2i+1$. The vertices are labeled by fractions of neighborhoods originated by 00, 01, 10 and 11, the overlap determines each connection. In Table~\ref{traslapes-generico} the intersections derived from the elements of each vertex are showed; they are the edges of the de Bruijn diagram as we can see in Figure~\ref{deBruijnGenerico}.

\begin{table}[th]
\centering
\small
\begin{tabular}{l|l}
(0,{\bf 0}) $\diamond$ ({\bf 0},0) & 0{\bf 0}0 \\
(0,{\bf 0}) $\diamond$ ({\bf 0},1) & 0{\bf 0}1 \\
(0,{\bf 1}) $\diamond$ ({\bf 1},0) & 0{\bf 1}0 \\
(0,{\bf 1}) $\diamond$ ({\bf 1},1) & 0{\bf 1}1 \\
(1,{\bf 0}) $\diamond$ ({\bf 0},0) & 1{\bf 0}0 \\
(1,{\bf 0}) $\diamond$ ({\bf 0},1) & 1{\bf 0}1 \\
(1,{\bf 1}) $\diamond$ ({\bf 1},0) & 1{\bf 1}0 \\
(1,{\bf 1}) $\diamond$ ({\bf 1},1) & 1{\bf 1}1
\end{tabular}
\caption{Intersections determining the edges of the de Brujin diagram.}
\label{traslapes-generico}
\end{table}

The de Bruijn diagram has four vertices which can be renamed as $\{0,1,2,3\}$ corresponding with the four partial neighborhoods of two cells $\{00,01,10,11\}$, and eight edges representing neighborhoods of size $2r+1$.

Paths in the de Bruijn diagram may represent chains, configurations or classes of configurations in the evolution space. 

\begin{figure}[th]
\centerline{\includegraphics[width=3.8in]{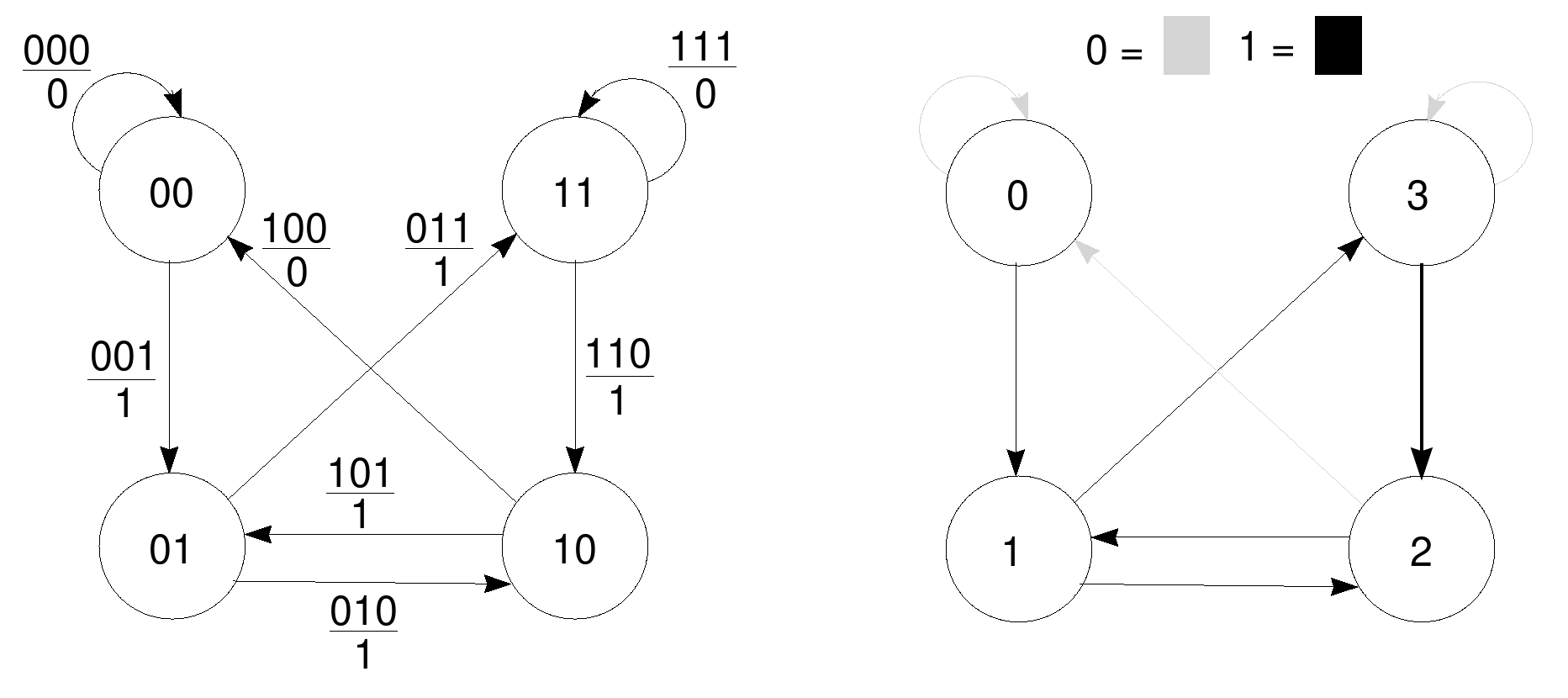}}
\caption{De Bruijn diagram for Rule 110.}
\label{deBruijnR110}
\end{figure}

The vertices of the de Bruijn diagram are sequences of symbols in the set of states and the symbols are sequences of vertices in the diagram. The edges describe how such a sequences can be overlapped; consequently, different intersection degrees produce distinct de Bruijn diagrams. Thus, the connection takes place between an initial symbol, the overlapping symbols and a terminal one (Table~\ref{traslapes-generico}).

De Bruijn diagram for Rule 110 is derived from the generic one (Figure~\ref{deBruijnGenerico}) and it is calculated in Figure~\ref{deBruijnR110}. The edge color represents the state in which each neighborhood evolves, as the second diagram of the same figure illustrates.

Now we must discuss another variant where the de Bruijn diagram can be extended to determine greater sequences by the period and the shift of their cells in the evolution space in Rule 110. A problem is that the calculation of extended de Bruijn diagrams grows exponentially with order $k^{{2r}^{n}}$ $\forall$ $n \in \mathbb Z^{+}$.

\begin{figure}[th]
\centerline{\includegraphics[width=4.8in]{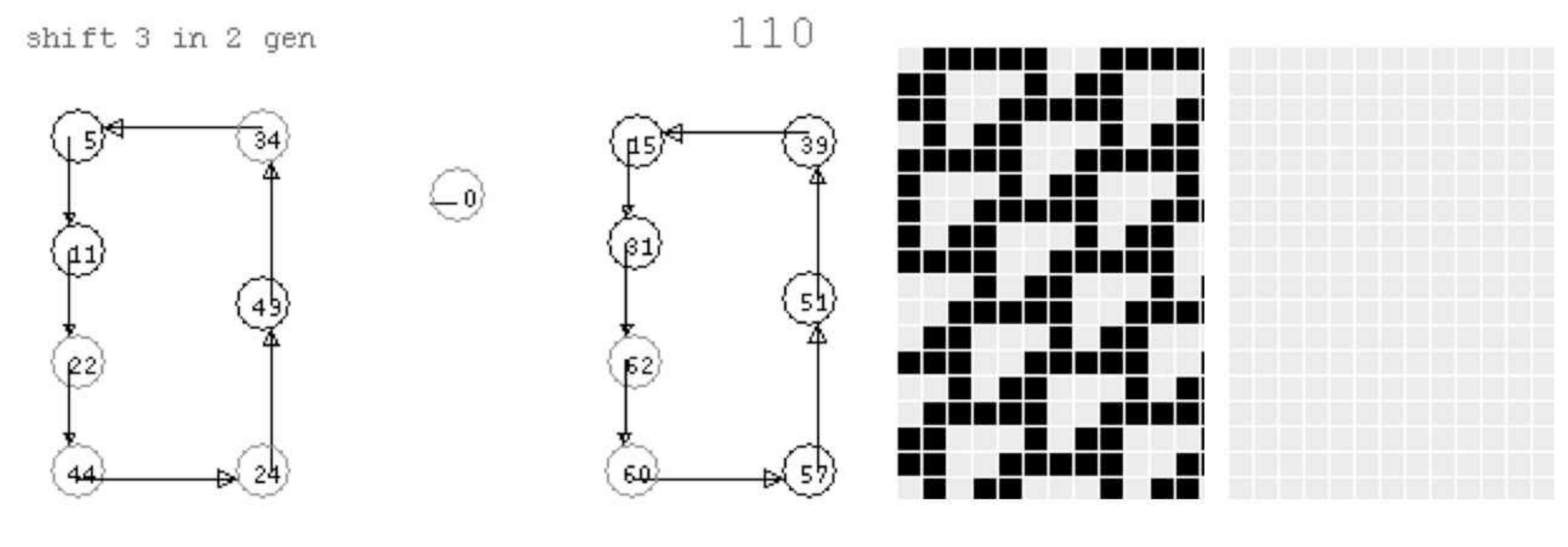}}
\caption{Extended de Bruijn diagram determining tiles: $T_0$ and  $T_3^{\alpha}$.}
\label{deBruijnExtendido}
\end{figure}

An extended de Bruijn diagram is illustrated in Figure~\ref{deBruijnExtendido}. The graphs of the left show the cycles in the diagram (at the right there are their respective evolutions). That means that not all the vertices offer relevant information; in fact we are only interested in the vertices forming cycles, because they determine periodic sequences following a particular path in the diagram. Figure shows three cycles; the first evolution illustrates the behavior of chains $(1100010)^*$ or $(1100111)^*$ determined by cycles of length 7, where the state is represented by the color of the vertex (for example, the vertex 5 (000101) intersect with vertex 11 (001011) forming the neighborhood 11 (0001011), that evolves into state 1). In this case, both cycles produce $T_{3}^{\alpha}$ mosaic with different chains. Also, the chains move to three elements to the right each two generations.

The behavior for the third cycle represented by vertex 0 produces all the sequences $0^+$. Second evolution of Figure~\ref{deBruijnExtendido} describes the behavior of this sequence dominated by tile $T_0$ (homogenous evolution).

The extended de Bruijn diagrams\footnote{The de Bruijn diagrams were calculated with the NXLCAU21 system developed by McIntosh for NextStep (OpenStep and LCAU21 to MsDos). Application and code source are available from: \url{http://delta.cs.cinvestav.mx/~mcintosh/oldweb/software.html}} calculate all the periodic sequences by the cycles defined in the diagram. These ones also calculate the shift of a periodic sequence for a certain number of steps; thus we can get de Bruijn diagrams describing all the periodic sequences characterizing a glider in Rule 110.

In order to explain how the sequences of each glider are determined, we firstly calculate the de Bruijn diagram composing an $A$ glider in Rule 110, and discussing how the periodic sequences are extracted for representing this glider and specifying as well the set of regular expressions.

\begin{figure}[th]
\centerline{\includegraphics[width=4.8in]{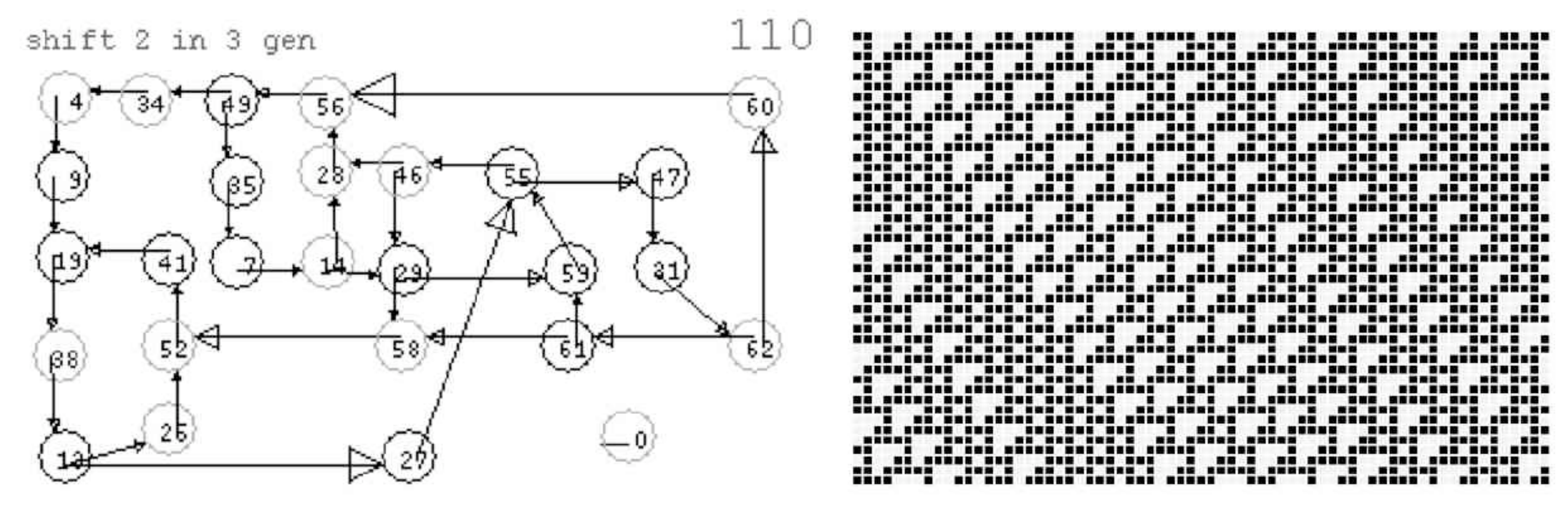}}
\caption{De Bruijn diagram calculating $A$ gliders and ether configurations.}
\label{fases-gliderA}
\end{figure}

The $A$ glider moves two cells to the right in three times (Table~\ref{margenesgliders}). We compute the extended de Bruijn diagram (2-shift, 3-gen) depicted in Figure~\ref{fases-gliderA}. The cycles of the diagram have the periodic sequences describing the $A$ glider; however, these sequences are not ordered yet. Therefore, we must determine and classify them.

In the figure we have two cycles: a cycle formed by vertex 0 and a large cycle of 26 vertices which is composed as well by 9 internal cycles. The evolution of the right illustrates the location of the different periodic sequences producing the $A$ glider in distinct numbers.

Following the paths through the edges we obtain the sequences or regular expressions determining the phases of the $A$ glider. For example, we have cycles formed by:

\begin{enumerate}
\item[I.] The expression (1110)*, vertices 29, 59, 55, 46 determining $A^{n}$ gliders.
\item[II.] The expression (111110)*, vertices 61, 59, 55, 47, 31, 62  defining $nA$ gliders with a $T_{3}$ tile between each glider.
\item[III.] The expression (11111000100110)*, vertices 13, 27, 55, 47, 31, 62, 60, 56, 49, 34, 4, 9, 19, 38 describing ether configurations in a phase (in the following subsection we will see that it corresponds to the phase $e$(f$_{1}$\_1)).
\end{enumerate}

The cycle with period 1 represented by vertex 0 produces a homogenous evolution with state 0. The evolution of the right (Figure~\ref{fases-gliderA}) shows different packages of $A$ gliders, the initial condition is constructed following some of the seven possible cycles of the de Bruijn diagram or several of them. We can select the number of $A$ gliders or the number of intermediate tiles $T_{3}^{\beta}$ changing from one cycle to another.

A problem on computing de Bruijn diagrams for all the periodic sequences representing each glider in Rule 110 is that the NXLCAU21 system is only able to estimate extended de Bruijn diagrams up to ten generations (implying an enormous diagram with 1,048,576 vertices); consequently, trying to order or classify all the cycles is a huge task. Also, as we can see in Table~\ref{margenesgliders}, $E$, $\bar{E}$, $F$, $G$, $H$ gliders and glider guns exceed by several times the limit of ten generations. In order to solve this problem and to determine all the regular expressions to each glider of Rule 110, we evaluate all the phases to each glider aligning tiles $T_3^{\beta}$. 

\begin{figure}[th]
\centerline{\includegraphics[width=6.3in]{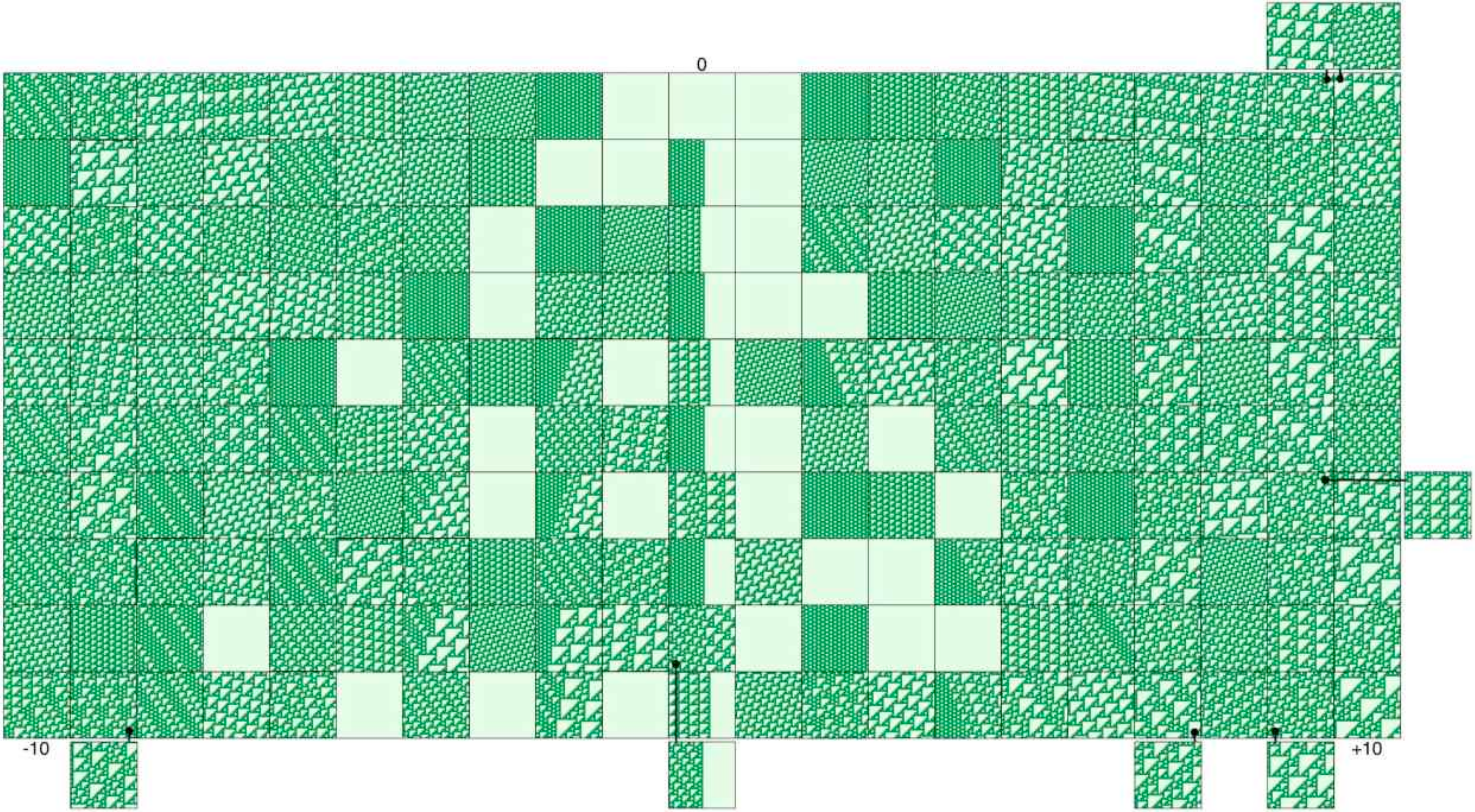}}
\caption{Patterns calculated by de Bruijn diagrams up to 10 generations.}
\label{deBruijn10genSmall}
\end{figure}

Let us take all the existing patterns derived from the de Bruijn diagrams (Figure~\ref{deBruijn10genSmall}) up to 10 generations and analyze some results briefly, an extensive discussion  can be found in \cite{kn:JMS06}. When the two numbers coincide the diagram consists exclusively of loops, but not necessarily of one single loop. Since zero is a quiescent state, entries of form (1,1) indicate that it is the only configuration holding the shifting requirement; in particular, there are no still life patterns (except for zero).

Some interesting points of the figure are that some of Cook's gliders are at entries (2,3) ($A$-gliders), (-2,4) ($B$-gliders), (0,7) ($C$-gliders), and at (2,10) ($D$-gliders). Notation $(x,y)$ indicates a shift of $x$ places, (negative values corresponding with a left shift) in $y$ generations.

Cook's gliders are found in different phases. For example at (-2,3) the $A$ glider completely covers the evolution space, at (-6,2) a package of $A^{2}$ gliders is interchanged with a $T_{3}$ tile, at (-10,1) there are $A^{3}$ gliders, at (-6,6) there are $A^{4}$ gliders, at (-6,8) there are $A^{5}$ gliders, at (-8,7) there are $A^{6}$ gliders, at (-8,9) there are $A^{7}$ gliders and so on. But also we can see configurations grouping $T_{3}$ tiles in different package of $A$ gliders as it can be seen at (-8,6), (-8,10), (-10,8), (4,6) and (6,9).

Another important point is that de Bruijn diagrams can find periodic configurations constructed by large tiles. For example in coordinate (10,10) we have that a $T_{11}$ tile may cover the evolution space with other additional tiles; we can find similar evolutions for tiles $T_{10}$, $T_{9}$, $T_{8}$, $T_{7}$ among others. The construction of the de Bruijn diagrams allows to validate each of the strings representing every glider of $\Psi_{R110}$, and we can apply well-known results from theory of languages like the pumping lemma or decision algorithms \cite{kn:HU79}.

The subset diagram \cite{kn:Mc91} is derived from the de Bruijn diagram, representing a general diagram for determining what sequences belong to the language produced by Rule 110 and besides defining the configurations in the Garden of Eden (sequences with no ancestors).

In this way, the subset diagram has $2^{k^{2r}}$ vertices, if all the configurations of certain length have ancestors then all the configurations with extensions both to the left and the right with the same equivalence must have ancestors. If this is not the case, then they describe configurations in the Garden of Eden and represent paths going from the maximum set to the minimum one in the subset diagram.

The nodes are grouped into subsets, note being taken of the subsets to which one can arrive through systematic departures from all the nodes in any given subset. The result is a new graph, with subsets for nodes and links summarizing all the places that one can get to from all the different combinations of starting points. Sometimes, but far from always, the possible destinations narrow down as one goes along; in any event one has all the possibilities cataloged.

One point to be observed is that if one thinks that there should be a link at a certain node and there is not, the link should be drawn to the empty set instead; a convention which assures every label of having a representation at every node in the subset diagram.

Vertices of the subset diagram are formed by the combination of each subset formed from the states forming the de Bruijn diagram (a power set). For example for a CA $(2,1)$ we have four sequences of states in the Bruijn diagram enumerated as $\{0\}$, $\{1\}$, $\{2\}$ and $\{3\}$, all the possible subsets are: $\{0,\ 1,\ 2,\ 3\}$, $\{0,\ 1,\ 2\}$, $\{0,\ 1,\ 3\}$, $\{0,\ 2,\ 3\}$, $\{1,\ 3,\ 2\}$, $\{0,\ 1\}$, $\{0,\ 2\}$, $\{0,\ 3\}$, $\{1,\ 2\}$, $\{1,\ 3\}$, $\{3,\ 2\}$, $\{3\}$, $\{2\}$, $\{1\}$, $\{0\}$ and $\{\}$. In these subsets four unitary classes can be distinguish; the incorporation of the empty set guarantees that all subsets have at least one image, although this one does not exist in the original diagram. In order to determine the type of union between the subsets, the state in which each sequence evolves must be reviewed to know towards which states (subset that form it) may be connected; this way the relation for Rule 110 is constructed in Table~\ref{relationsubsetR110}.

There is another important reason for working with subsets. Labelled links resemble functions, by associating things with one another. But if two links with the same label emerge from a single vertex, they can hardly represent a function. Forging the subset of all destinations, leaves one single link between subsets, bringing functionality to the subset diagram even though it did not exist originally. Including the null set ensures that every point has an image, avoiding partially defined functions.

Once the subset diagram has been formed, if a path leads from the universal set to the empty set, that is conclusive evidence that such a path exists nowhere in the original diagram. Another applicationÑthe one originally envisioned by Edward Moore \cite{kn:Moore56}--is to determine whether there are paths leading to the unit classes. Such a paths, if they existed, could be used to force an automaton into a predetermined state, no matter what its original condition

\begin{table}[th]
\centering
\begin{tabular}{|c|c|c|} 
\hline
vertex & edge with 0 & edge with 1 \\
\hline
$0$ & $0$ & $1$ \\
$1$ & $\phi$ & $2,3$ \\
$2$ & $0$ & $1$ \\
$3$ & $3$ & $2$ \\
\hline
\end{tabular}
\caption{Relation between states of the subset diagram.}
\label{relationsubsetR110}
\end{table}

Although the edges between subsets do not define a function, it is well defined for the whole graph by the inclusion of the empty set. Each class of edges defines a function: $\Sigma_{0}$ or $\Sigma_{1}$. The subset diagram describes the join of $\Sigma_{0} \cup \Sigma_{1}$, that by itself is not functional.

Let $a$ and $b$ be vertices, $S$ a subset and $|S|$ the cardinality of $S$; then the subset diagram is defined by the following equation:

\begin{equation}
	\sum_{i} (S) = \left\{\begin{array}{lll}
				\phi & S=\phi \\ 
       		                  \{b\ |\ \mbox{edge}_{i}\ (a,b)\} & S=\{a\}. \\
				\bigcup_{a \in S} \Sigma_{i}(a) & |S|>1
		         \end{array} \right. 
\end{equation}

\noindent three important properties are given here:

\begin{enumerate}
\item If there is a path from the maximum subset to the minimum one, then there exists a similar path starting from some smaller subset to the empty one. On the other hand, if all the unitary classes do not have edges going to the empty set, then there are no configurations in the Garden of Eden. 
\item There is a certain image of the de Bruijn diagram, in the sense that given an origin and a destiny, there is always a subset containing the accessible destiny and another subset containing the origin, besides the destiny can have additional vertices.
\item The subset diagram is not connected, and it is interesting to know the accessible greatest subset as well as the smallest one from a given subset.
\end{enumerate}

The local function $\varphi$ of Rule 110 has an injective correspondence, knowing this correspondence then we must find paths in the subset diagram going from the maximum set to the empty set. Two minimal configurations in the Garden of Eden of Rule 110 are: (101010)* and (01010)*.

\begin{figure}[th]
\centerline{\includegraphics[width=1.2in]{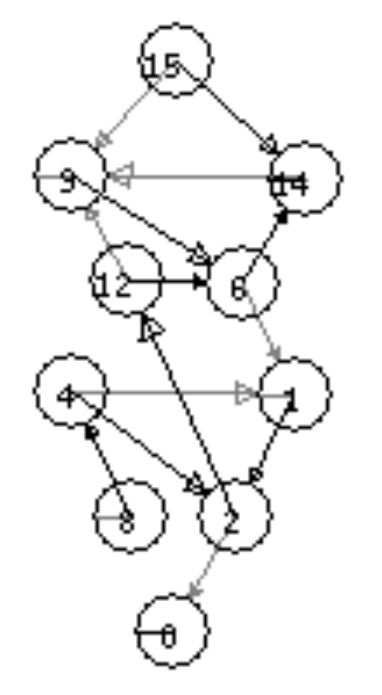}}
\caption{Subset diagram of Rule 110.}
\label{subset-diagrama}
\end{figure}

Also of obtaining the Garden of Eden sequences through the subset diagram. We have too a general machine recognizing each sequence in $\Psi_{R110}$. In order to verify this it is just necessary to take a sequence from the subset of regular expressions, hence there exists a path in the subset diagram starting from the maximum set determining its existence on ending into a nonempty subset.

Altogether, the principal value of the scalar subset diagram is to establish such things as:

\begin{enumerate}
\item The shortest excluded words, the occurrence of any one of which creates a Garden of Eden configuration.
\item A maximum length for a minimal excluded word, which is the number of nodes in the portion of the subset diagram connected to the full subset.
\item Whether exclusion occurs in stages, as key segments are built up.
\item A regular expression describing excluded words.
\end{enumerate}

\subsection{Phases in Rule 110}

In this section we discuss how the phases are derived, represented and obtained to determine periodic sequences in the evolution space of Rule 110.

\begin{figure}[th]
\centerline{\includegraphics[width=1.2in]{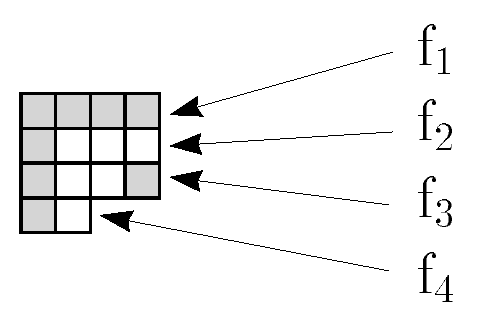}}
\caption{Phases f$_{i}$ of the $T_{3}$ tile.}
\label{T3}
\end{figure}

The $T_{3}^{\beta}$ tile illustrated in Figure~\ref{T3} has four phases or sequences by row: f$_{1}$ = 1111, f$_{2}$ = 1000, f$_{3}$ = 1001, and f$_{4}$ = 10 (from now on we shall simply talk about $T_{3}^{\beta}$ tile as $T_{3}$). Thus, the concatenation of four phases f$_{i}$ determine a (periodic) sequence describing the ether pattern: f$_{1}$f$_{2}$f$_{3}$f$_{4}$ = 11111000100110.

Following each level of $T_{3}$ we determine that there are at most four phases to represent any periodic sequence. First we derive all the possible phases of ether in Rule 110 and define them in the following way: $e$(f$_{1}$\_1) = 11111000100110, $e$(f$_{1}$\_2) = 10001001101111, $e$(f$_{1}$\_3) = 10011011111000, and $e$(f$_{1}$\_4) = 1011111000\-1001.

The evolution of Rule 110 converges in time generally into ether from random initial conditions with a 0.57 of probability. The initial condition constructed by the expression $e$(f$_{1}$\_$i$)* $\forall$ $1 \leq i \leq 4$, where the interval indicates all the possible phases, covers the whole evolution space with ether. Let us notice that each  phase $e$(f$_{1}$\_$i$) is a permutation of first one. Therefore, fixing a phase is sufficient to establish a measurement; by sequential order we chose phases f$_{i}$\_1 to establish a horizontal one.

\begin{figure}[th]
\centerline{\includegraphics[width=4.8in]{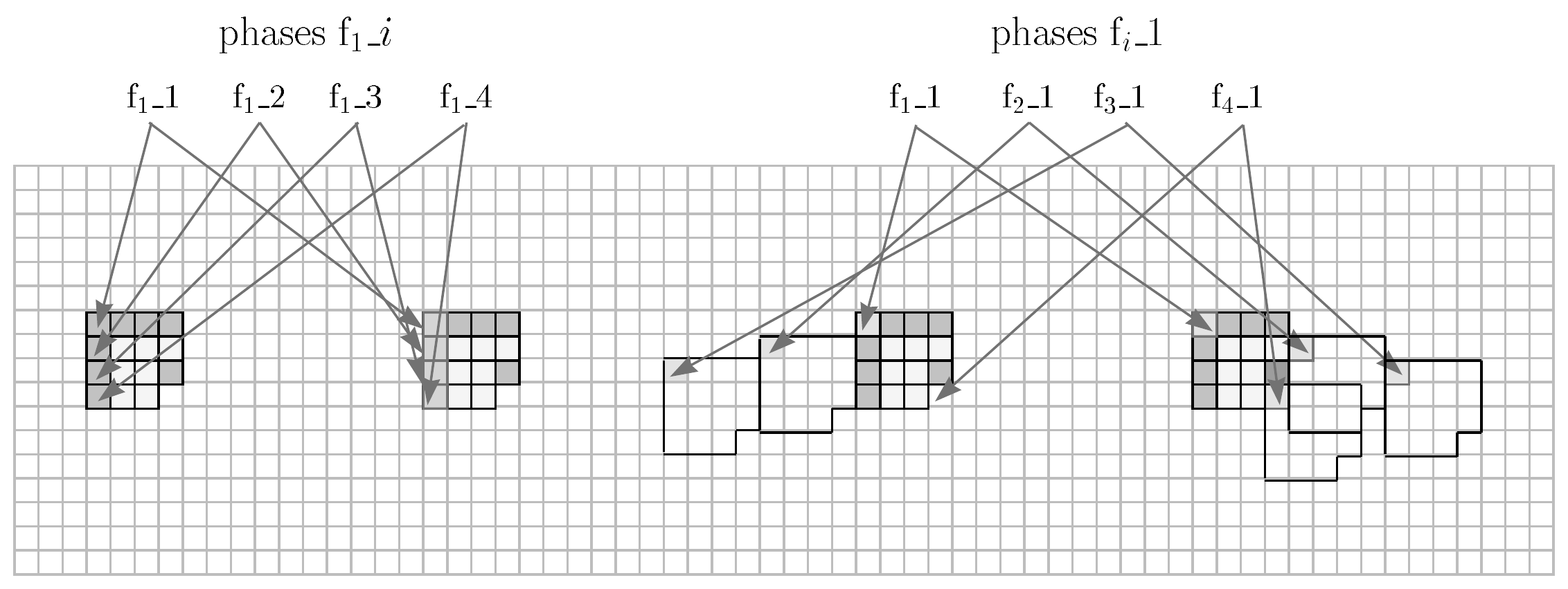}}
\caption{Phases f$_{i}$\_$i$ with the $T_{3}$ tile.}
\label{fases}
\end{figure}

Cook determines two measures in the evolution space \cite{kn:Cook04}: horizontal $\frown^i$ and vertical $\nearrow^i$. We only determine the horizontal case f$_{i}$\_1. Phases f$_{i}$\_1 have four sub-levels consequence of the phases in $T_{3}$ tile (Figure~\ref{fases}, left part) and each phase can be aligned $i$ times generating all the possible phases (right part).

The phases represent the periodic sequences (regular expressions of each glider) of finite length in the de Bruijn diagram. It is important to indicate that an alignment of a phase determines a set of regular expressions and another alignment defines another set of them. Thus, we have four possible sets (Table~\ref{fases-fi_i}): $Ph_{1}$ (phases level one), $Ph_{2}$ (phases level two), $Ph_{3}$ (phases level three) and $Ph_{4}$ (phases level four), where the sets are disjunct each other to construct initial conditions. The property of regular expressions is conserved only in the domain of each set if we want to project these structures in the dynamics of the cellular automaton, where the separation is originated by the four permutations describing ether. In this way there are four sets where the elements of one  are permutations the elements in other; therefore a single set is enough to construct initial conditions under the rules of regular expressions.

The way of calculating the whole set of strings for every glider is analyzing the alignment of the f$_{i}$\_1 phases. In order to determine them, first it is necessary to describe each glider in its form and limits through tiles. Later we fix a phase, in our case we took f$_{i}$\_1 and we drew up to a horizontal line in the evolution space tying two tiles $T_{3}$ (second illustration in Figure~\ref{fases}). Thus, the sequence between both tiles aligned in each one of the four levels determines a periodic sequence representing a particular structure in the evolution space of Rule 110. We calculate all the periodic sequences in a certain phase and this procedure enumerates all the periodic sequences forming each glider.

\begin{table}[th]
\centering
\begin{tabular}{rcc}
phases level one ($Ph_1$) & $\rightarrow$ & $\{$f$_{1}$\_1, f$_{2}$\_1, f$_{3}$\_1, f$_{4}$\_1$\}$ \\
phases level two ($Ph_2$) & $\rightarrow$ & $\{$f$_{1}$\_2, f$_{2}$\_2, f$_{3}$\_2, f$_{4}$\_2$\}$ \\
phases level three ($Ph_3$) & $\rightarrow$ & $\{$f$_{1}$\_3, f$_{2}$\_3, f$_{3}$\_3, f$_{4}$\_3$\}$ \\
phases level four ($Ph_4$) & $\rightarrow$ & $\{$f$_{1}$\_4, f$_{2}$\_4, f$_{3}$\_4, f$_{4}$\_4$\}$
\end{tabular}
\caption{Four sets of phases $Ph_i$ in Rule 110.}
\label{fases-fi_i}
\end{table}

Variable f$_{i}$ indicates the phase currently used where the second subscript $i$ (forming notation f$_{i}$\_i) indicates that selected set $Ph_i$ of regular expressions. Finally, our notation proposes to codify initial conditions by phases is in the following way:

\begin{equation}
\#_{1}(\#_{2},\mbox{f}_{i}\_1)
\end{equation}

\noindent where \#$_{1}$ represents the glider according to Cook's classification (Table~\ref{margenesgliders}) and \#$_{2}$ the phase of the glider if it has a period greater than four.\footnote{We must indicate that the arrangement by capital letters for the \#$_{2}$ parameter into the OSXLCAU21 system \cite{kn:Jua04} does not have a particular meaning; it is only used to give a representation at the different levels for phases with gliders of periods module four.}

\begin{figure}[th]
\centerline{\includegraphics[width=5in]{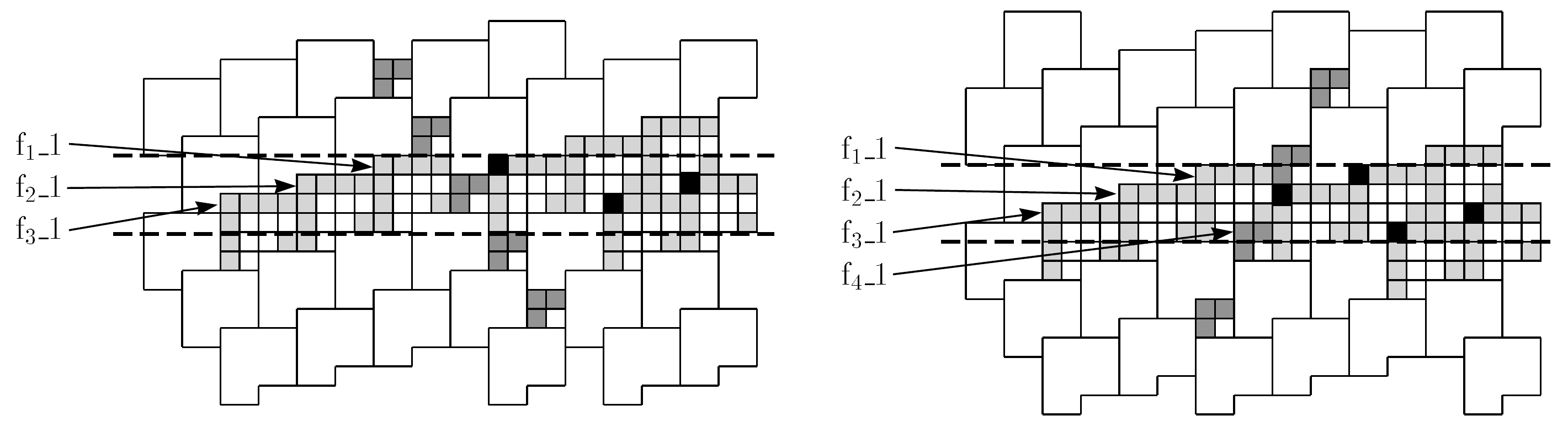}}
\caption{Phases f$_{i}$\_1 for $A$ and $B$ gliders respectively.}
\label{fases-gliderA-2}
\end{figure}

Now we determine the phases f$_{i}$\_1\footnote{The subset of regular expressions $\Psi_{R110}$ for each glider in Rule 110 (see Appendix), serves as input data for the OSXLCAU21 system \cite{kn:Jua04}.} for $A$ and $B$ gliders as Figure~\ref{fases-gliderA-2} illustrates. $T_{3}$ tiles determine a phase $\#_{1}$;  in the case of $A$ and $B$ gliders only a $T_{3}$ tile is necessary to describe their structure. In all the others cases, at least two $T_{3}$ tiles are needed.

Following each phase initiated by every $T_{3}$ tile, the phases f$_{i}$\_1 for the $A$ glider are as follows:

\begin{itemize}
\item $A$(f$_{1}$\_1) = 111110
\item $A$(f$_{2}$\_1) = 11111000111000100110
\item $A$(f$_{3}$\_1) = 11111000100110100110
\end{itemize}

The sequence is defined taking the first value from the first cell of $T_{3}$ tile on the left until reaching a second cell representing the first value of the second $T_{3}$ tile on the right. In Figure~\ref{fases-gliderA-2} a black cell indicates the limit of each phase.

In general for every structure with negative speed, the phase f$_{4}$\_1 = f$_{1}$\_1, for this reason the phase is not written. Each periodic sequences defined by $T_{3}$ tiles conserves the regular expression property when basic rules are applied. Therefore, $\epsilon$, $A$(f$_{1}$\_1), $A$(f$_{1}$\_1)+$A$(f$_{1}$\_1), $A$(f$_{1}$\_1)-$A$(f$_{1}$\_1), $A$(f$_{1}$\_1)* and $A$(f$_{3}$\_1)-$A$(f$_{1}$\_1)-$A$(f$_{2}$\_1)-$A$(f$_{3}$\_1)-$A$(f$_{2}$\_1) are regular expressions (we use `-' to represent the concatenation operation in our constructions). Let us remember the codification in phases, $A$ indicates the glider ($\#_1$) and f$_i$\_1 indicates the phase.

Thus, all phases f$_{i}$\_1 for the $B$ glider are:

\begin{itemize}
\item $B$(f$_{1}$\_1) = 11111010
\item $B$(f$_{2}$\_1) = 11111000
\item $B$(f$_{3}$\_1) = 1111100010011000100110
\item $B$(f$_{4}$\_1) = 11100110
\end{itemize}

The procedure made over $A$ and $B$ gliders was applied to all the gliders for obtaining the whole subset of regular expressions $\Psi_{R110}$. First we shall expose some properties of $T_3$ tile representing ether in Rule 110 and how they are reflected in the evolution space for each periodic and non-periodic structure.

\begin{figure}[th]
\centerline{\includegraphics[width=4.8in]{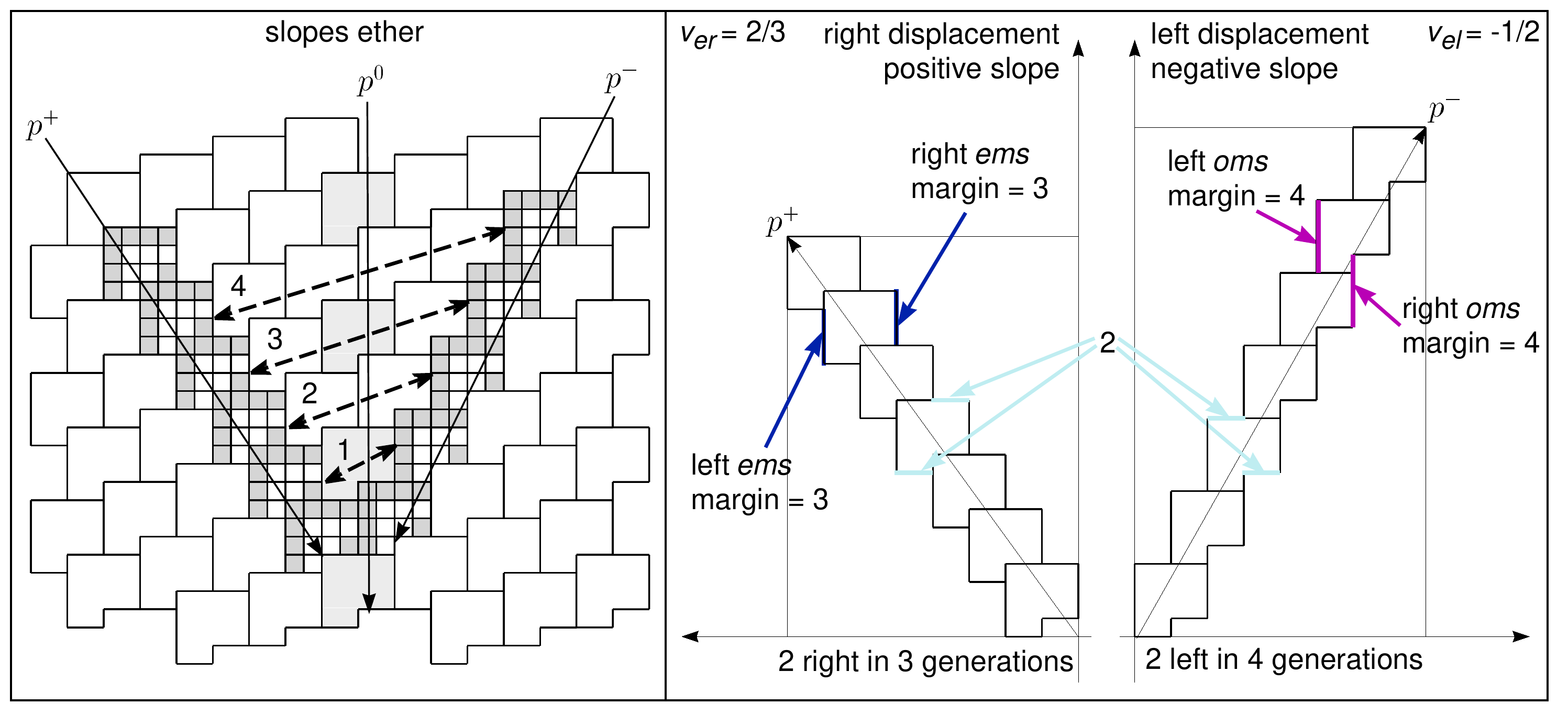}}
\caption{Three slopes produced by the ether pattern.}
\label{pendientes_ether}
\end{figure}

The $T_{3}$ tile determines three types of slopes\footnote{If $P_{1}(x_{1},y_{1})$ and $P_{2}(x_{2},y_{2})$ are two different points one a straight line, its slope $m$ is: $m = \frac{y_{1}-y_{2}}{x_{1}-x_{2}}$. Thus we can select a first point $(i,j)$ into the evolution space of Rule 110 within some $T_{3}$ tile for each one of its shifts. If the shift goes from left to right, the second point is $(i+2,j+3)$. If the shift goes from right to left, the second point is $(i-2,j+4)$.} as we can see in Figure~\ref{pendientes_ether}: positive slope ``$p^{+}$,'' negative slope ``$p^{-}$'' and null slope ``$p^{0}$.'' The slopes $p^{+}$ and $p^{-}$ specify maximal positive and negative speeds for all the gliders in the evolution space of Rule 110.

If $p^{+}$ has a shift of +2 cells in 3 generations, then the ether speed is $v_{er}=2/3$. If $p^{-}$ has a shift of -2 cells in 4 generations, then the speed of the ether is $v_{el}=-1/2$ (as we have indicated in Table~\ref{margenesgliders}).

In the analysis by phases the $T_{3}$ tile determines the existence of two margins ``$oms$'' and ``$ems$'' (right illustration in Figure~\ref{pendientes_ether}) for both slopes and each tile, establishing other important properties.

If $p^{+}$, there is an odd margin $oms$ with a height of three cells. If $p^{-}$, there is an even margin $ems$ with a height of four cells. The contact points\footnote{A contact point \cite{kn:PST86} indicate as a region where a given glider may hit against another one. Therefore, a non-contact point implies a region where a glider cannot be affected by some collision. Contact points are not exclusive for periodic structures in Rule 110, but they also exist in non-periodic structures.} are determined by the number of odd margins $oms$ when $p^{+}$ or even margins $ems$ when $p^{-}$. Finally, both odd and even margins (left and right in a periodic structure) have a bijective correspondence (see Table~\ref{margenesgliders}).

If there are $n$ margins $oms$ in the upper part of a glider when $p^{+}$, then there are $n$ margins $oms$ in its lower part. In the other hand, if there are $n$ margins $ems$ in the upper part when $p^{-}$, then there are $n$ margins $ems$ in its lower part. In other words, the existence of a contact point in a glider implies the existence of a non-contact point in its converse part. 

All periodic or non-periodic structure in the evolution space of Rule 110 advances $+2$ cells and goes back $-2$ cells, then each structure with $p^{+}$ has a speed of $v_g \leq v_{er}$ and when $p^{-}$ then $v_g \leq |v_{el}|$, where $v_g$ represents the speed of a $g$ glider (see Table~\ref{margenesgliders}). Therefore, every structure with $p^{+}$ advances with increments $v_{er}$ and goes backs with decrements $v_{el}$. In other case, the structure with $p^{-}$ advances with increments $v_{el}$ and goes back with decrements $v_{er}$. 

Every structure with $p^{+}$ can be affected by another structure with different slope ($p^{0}$ or $p^{-}$), only if the first has at least a margin $oms$ and the second has at least a margin $ems$ . In the other case, each structure with $p^{-}$ can be affected by another structure with distinct slope ($p^{0}$ or $p^{+}$), only if the first has at least a margin $ems$ and the second has at least a margin $oms$.

These properties produce the following equations. Let $\mathcal G$ be the whole set of gliders in Rule 110, then the shift of $g \in \mathcal G$ is represented in the following way:

\begin{equation}
d_{g}=(2*oms)-(2*ems).
\label{eq-desplazamiento}
\end{equation}

Every periodic structure has a period defined by the number of margins $oms$ and $ems$. Therefore, the period of a $g$ glider is determined by:

\begin{equation}
p_{g}=(3*oms)+(4*ems)
\label{eq-periodo}
\end{equation}

\noindent and has a speed described by:

\begin{equation}
v_{g} = \frac{(2*oms)-(2*ems)}{(3*oms)+(4*ems)}.
\label{eq-velocidad}
\end{equation}

The number of collisions between gliders have a maximum level determined by the number of margins $oms$ and $ems$. Thus, for an arbitrary glider with $oms$ contact points and other arbitrary glider different from the first with $ems$ contact points, we have the following number of collisions:

\begin{equation}
c \leq oms * ems
\label{eq-choquesmax}
\end{equation}

\noindent where $c$ represents the maximum number of collisions between both gliders. Nevertheless, in some gliders the maximum level is not fulfilled. Depurating the equality we have exact number of collisions between a pair $g_{i}$, $g_{j} \in \mathcal G$ where $i \neq j$ in the following equation: 

\begin{equation}
c = |(oms_{g_{i}} * ems_{g_{j}}) - (oms_{g_{j}} * ems_{g_{i}})|.
\label{eq-choques}
\end{equation}

The procedure used to codify initial conditions by phases f$_{i}$\_1 to handle collisions, specifies as well two measures representing distances in the linear space of Rule 110 (Table~\ref{distancias-fi_1}): mod 4 (by each $T_{3}$ tile). In this case we have a minimum distance of zero $T_{3}$ tiles and a maximum distance of three $T_{3}$ tiles among gliders. The second measurement is module 14 (by number of cells). In this case we have a minimum distance of 0 cells up to $4+4+4+2$ cells among gliders. The restriction is generated by the $T_{3}$ tile.

\begin{table}[th]
\centering
\footnotesize
\begin{tabular}{|c||c||c|}
\hline
$p^{+}$ & $p^{-}$ & $p^{0}$ \\
\hline \hline
f$_{1}$\_1 $\mapsto$ 1$T_{3}$(right) - 0$T_{3}$(left) & f$_{1}$\_1 $\mapsto$ 1$T_{3}$(left) - 1$T_{3}$(right) & f$_{1}$\_1 $\mapsto$ 1$T_{3}$(left) - 0$T_{3}$(right) \\
f$_{2}$\_1 $\mapsto$ 2$T_{3}$(right) - 3$T_{3}$(left) & f$_{2}$\_1 $\mapsto$ 2$T_{3}$(left) - 0$T_{3}$(right) & f$_{2}$\_1 $\mapsto$ 2$T_{3}$(left) - 3$T_{3}$(right) \\
f$_{3}$\_1 $\mapsto$ 3$T_{3}$(right) - 2$T_{3}$(left) & f$_{3}$\_1 $\mapsto$ 3$T_{3}$(left) - 3$T_{3}$(right) & f$_{3}$\_1 $\mapsto$ 3$T_{3}$(left) - 2$T_{3}$(right) \\
f$_{4}$\_1 = f$_{1}$\_1 & f$_{4}$\_1 $\mapsto$ 0$T_{3}$(left) - 2$T_{3}$(right) & f$_{4}$\_1 $\mapsto$ 0$T_{3}$(left) - 1$T_{3}$(right) \\
\hline
\end{tabular}
\caption{Phases determining distances mod 4 (by $T_{3}$ tiles).}
\label{distancias-fi_1}
\end{table}

The relevance of knowing and determining a distance in the linear space of Rule 110 is for establishing a suitable control for the positions of gliders and for obtaining the desired reactions. Now we analyze the distances induced by the phases (by tile) for slopes $p^{+}$, $p^{-}$ and $p^{0}$, as described in Table~\ref{distancias-fi_1}.

In the case of the ether sequences, the distances are the same implying an interval of 4$T_{3}$ tiles which is the maximum distance by tile. For $p^{+}$ we can take an $A$ glider as example; for $p^{-}$ a $B$ glider can be chosen and for $p^{0}$ a $C$ glider is useful to verify the distances.

In this way, if a glider has a slope $p^{-}$ then the phases do not overlap, conversely if a glider has a slope $p^{+}$ the phases overlap. Finally, If the phase f$_{4}$\_1 overlaps with the phase f$_{1}$\_1 then in this case only we have three phases, in other one, we have four different periodic chains. Therefore, if f$_{4}$ \_1 overlaps with f$_{1}$\_1, f$_{4}$\_1 = f$_{1}$\_1.

If a glider with $p^{+}$ collisions sequentially in its four phases against a glider with $p^{-}$ in its four phases as well placed in $(x,y)$, then the collisions take place in the same $y$-position with identical distances between each interval. The minimum range is important to generate a proper collision. Kenneth Steiglitz determines two types of collisions \cite{kn:PST86} between gliders:

\begin{itemize}
\item {\it Proper collision}. The collision takes place in a contact point.
\item {\it Non-proper collision}. The collision takes place when the gliders overlap in their sequences, i. e., they overlap in initial conditions or in the evolution.
\end{itemize}

Rule 110 has the two forms of non-proper collisions: the first case is in phases overlapping from initial conditions and the second one is where several gliders interact in regions where distances cannot be extrapolated.

\subsection{Phases on non-periodic structures}

The phases not only can be determined in periodic structures of Rule 110, they are also defined in non-periodic structures.

The projection to non-periodic structures is made in the same way that we did with the periodic ones, only that the glider case is easier because they have a period and, therefore, a fixed number of phases. Nevertheless, for decompositions in short or long chaotic regions, the number of margins $oms$ and $ems$ varies arbitrarily. The chaotic regions can initiate from initial conditions or, in other cases, they are originated by collisions of two or more gliders and even by the near interaction of one or several chaotic regions.

\begin{figure}[th]
\centerline{\includegraphics[width=3.5in]{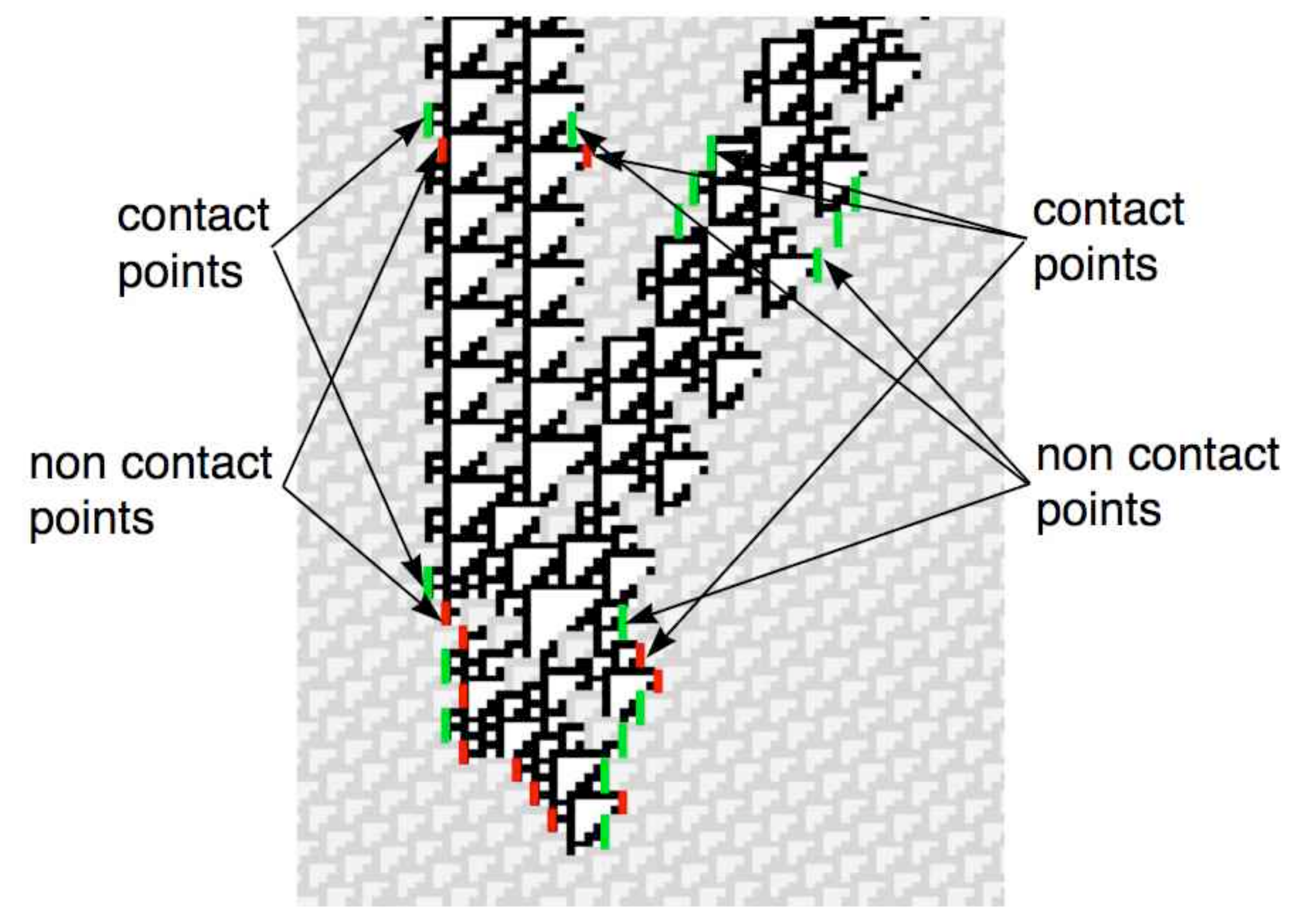}}
\caption{Annihilation of gliders with a short decomposition.}
\label{descomp}
\end{figure}

In Figure~\ref{descomp} we show a small decomposition generated by the collision among three gliders. The sequence for this example is: $e$*-$C_{2}$(A,f$_{3}$\_1)-$C_{2}$(A,f$_{1}$\_1)-$e$-$\bar{B}$(A,f$_{2}$\_1)-$e$*.\footnote{From now on the phase represented by an ether configuration `$e$(f$_1$)' will be simply described as `$e$' because it never changes.}

$C_{2}$ glider has both an $oms$ and an $ems$ margin in each end and $\bar{B}$ glider has three $oms$ and zero $ems$ margins in each end (see Figure~\ref{margenesgliders}). The chaotic region has eight $oms$ and three $ems$ margins in the left part, but in the right one it has three $oms$ and five $ems$ margins. In this case, the decomposition has several points where other structures can interact in both margins. Consequently, for a non-periodic structure the bijective correspondence between $oms$ and $ems$ margins is not conserved, because it does not have a specific period.

Concluding, every structure in the evolution space of Rule 110 must have at least a contact point and other non-contact point.

For example, in Figure~\ref{descomp} we can seen two $C_{2}$ gliders close to (2$C_{2}$) with $ems$ and $oms$ margins, conserving the pair of gliders without alteration. So, at the end of the chaotic decomposition, the last collision is between an $A$ glider against a $B$ glider with a distance of 0$T_{3}$ tiles. For this reason the decomposition does not leave debris at the end of its evolution. The product of the collision among $A$ and $B$ gliders with distance 0 can be specified with the following expression: $e$*-$A$(f$_{1}$\_1)-$B$(f$_{4}$\_1)-$e$*.

\subsection{A simple procedure to construct desired collisions}

The goal of the procedure is to construct initial conditions in the one-dimensional space of Rule 110 for controlling collisions in the evolution space. The constructions are codified by the phases f$_i$\_1 determining as well the base set of regular expressions f$_i$\_1; offering a procedure to handle complex collisions among all the possible structures.

Let $\Psi_{R110}$ be the base subset of regular expressions determined by the set of gliders $\mathcal G$. Now we specify a subset: $\epsilon$, 0, 1, $e$ and \#$_{1}$(\#$_{2}$,f$_{i}$\_1) $\in \Psi_{R110}$ as regular expressions following the classic rules. Thus we have two ways of yielding collisions among gliders:

\begin{itemize}
\item The first case is fixing the initial phases of two $g_{i}$, $g_{j}$ $\in \mathcal G$ where $i \neq j$ and both have different slopes $p^+$ and $p^-$ (or $p^0$). Then the interval between the two gliders is determined by an ether sequence $e$ and with this condition we can enumerate all the possible binary collisions just changing the interval of ether, i. e., manipulating the distance. Therefore, the collision between two gliders is determined by the expression: $e^+$-$g_{i}$-$e^*$-$g_{j}$-$e^+$. The restriction is that it only enumerates only collisions between two gliders and not among packages of them.

\item The second case can codify several equal or different gliders simultaneously, where the phase and distances are particularly manipulated, i. e., we can change both parameters to obtain a collision in the wished time and place. The advantage to use this case, is that we obtain a total control of the evolution space, but the disadvantage is that in order to determine the collision we must evaluate the production of the initial condition to know the distance and the adequate phase to get the desired result. In other words, we must construct the codification by a proof-and-error approach, but we will see that it is the best option because several collisions must be codified changing the phase, but not the distance (we exemplified this situation in the following section, when we applied the procedure to construct specific initial conditions for solving a particular problem).
\end{itemize}

For instance, if we want to produce a given glider by collisions among others, the involved speed of each glider may be different in every case. Therefore, if we need a simultaneous collision, we must determine first the distance and later the phase among them to obtain the required reaction in the wished place.

We present a number of steps to construct initial conditions in Rule 110 involving several gliders, we remark that the result shall be obtained by subsequent approaches.

\begin{enumerate}
\item Determine the number of $g_{i}$ gliders where $i \in \mathbb Z^{+}$ and the particular $g \in \mathcal G$ desired in the process.
\item Determine the f$_{i}$\_1 phase where $1 \leq i \leq 4$ in which each glider must start.
\item Determine the distance defined by ether $e$ between each glider (if it is necessary).
\item Execute the assigned codification to evaluate the production. If the production is correct, finish the allocation. In other case:
\begin{enumerate}
\item If the distance is correct but the phase is not the right one, a search is made crossing all $j$ phases of $\#_{2}$, where $1 \leq j \leq p_{g}$ and $p_{g}$ is the number of possible phases established by the number of margins $4ems + 3oms = p_{g}$ in a $g$ glider of period $p_{g}$.
\item If the phase is correct but the distance is not the right one, it is necessary to calculate the number of configurations $ne$ + \#$_{1}$(\#$_{2}$,f$_{i}$\_1) (mod 4 or mod 14). Establish if it is necessary to assign ether configurations or just change the phase of the structure.

If the distance is smaller to mod 4 (0, 1, 2 or 3$T_{3}$ tiles), it is not necessary to introduce a sequence $e$. In this case we need to adjust the distance with the $i$ phases f$_{i}$\_1 of $g$. If distance is mod 14 (4, 4, 4 or 2 number of cells), follow the previous criterion and return to step 4.
\end{enumerate}
\end{enumerate}

The complexity grows in the evolution space of Rule 110 with regard of the number of gliders involved and the size of the initial configuration, by the information amount contained in the chain.

\begin{figure}[th]
\centerline{\includegraphics[width=4.8in]{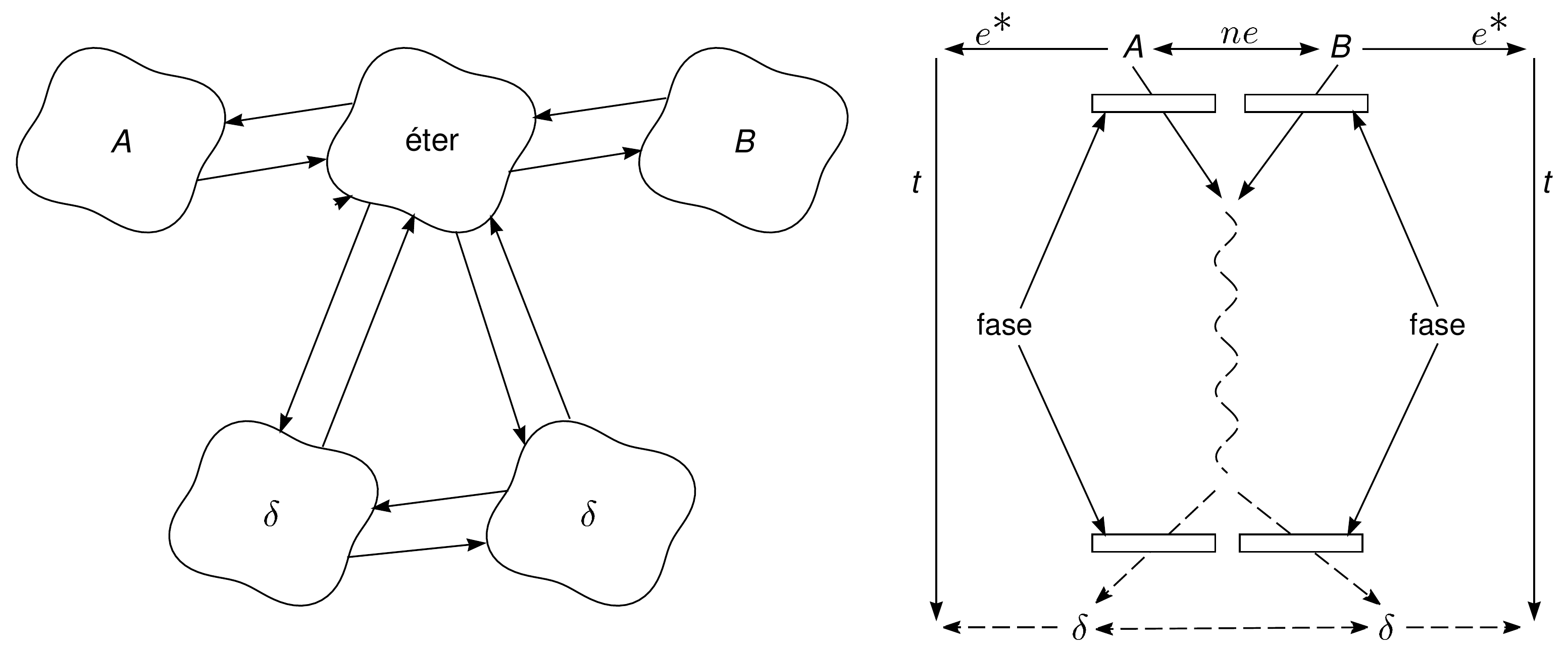}}
\caption{Schematic diagram representing collisions in 1D CA.}
\label{esqFaseChoque-2}
\end{figure}

All word $w$ constructed through phases under the basic rules of regular expressions represents an initial condition, in Figure~\ref{esqFaseChoque-2} we show the schematic diagram describing collisions and phases by periodic sequences in the evolution space.

The relation between two cycles in the de Bruijn diagram may be with two or more different periodic chains. In the figure we have that an $A$ glider has a connection with ether and on the other hand ether has a connection with a $B$ glider. The final sequence is an assigned regular expression in the initial condition yielding a collision. The result is interpreted in one or several $\delta$ gliders (we do not know the result of the reaction).

The right diagram of the figure represents exactly what does a phase mean. For example, we assign to the initial condition arbitrary ether sequences in both ends and a fixed phase between $A$ and $B$ gliders. Consequently, the periodic sequence between ether sequences is not altered in its length, and it will represent the glider phases during its movement through the evolution space.

\section{Conclusions}

The basic structure of Rule 110 has been explained, its behaviors and all the gliders until now known were displayed. The phases were described in detail, showing their origin induced by the analysis both in de Bruijn diagrams and tiles in Rule 110. Thus, phases help to determine a classification of periodic sequences to obtain the subset of glider-based regular expressions.

Also, once obtained the subset of glider-based regular expressions we proposed a codification by phases `$\#_1$($\#_2$,f$_i$\_1)' to construct initial conditions in the one-dimensional space of Rule 110 for determining a procedure to control collisions between gliders from initial conditions.

The application of this regular set has been used to describe to the universe of gliders in Rule 110 \cite{kn:JMS06} and the construction of Rule 110 objects \cite{kn:JMSa,kn:JM01} besides to other interesting reactions; for instance, the reconstruction of the operation of the cyclic tag system \cite{kn:JSM}.\footnote{ You can see a full snapshots and details description by components of functioning of cyclic tag system in Rule 110 from \url{http://uncomp.uwe.ac.uk/genaro/rule110/ctsRule110.html}}

Finally, several questions arise because it seems that the evolution of Rule 110 language should always be regular. For instance, How a regular language can be able of constructing a universal machine? Could Rule 110 determine new grammars? \cite{kn:MC00,kn:MC98,kn:Hurd87}. Could we project this language to two-dimensional finite-state automata? \cite{kn:KC01,kn:MC98}. Could Rule 110 be able of implementing unconventional logic operations by glider-based reactions? \cite{kn:Ada01,kn:Ada02}. Well, it is only the beginning.

\section*{\bf Acknowledgement}

\noindent This paper was inspired by the results of Prof. Harold V. McIntosh in de Bruijn diagrams and Rule 110. Prof. McIntosh has been as well an invaluable professor in M\'exico for a huge number of researches; in our case his work has been a major influence since our initial studies in cellular automata theory, in particular in the application of graph theory, algebra of matrices, CAMEX, probability and statistics, between other topics. Particularity, he was our first contact with CAM-PC and NXLCAU systems, stimulating in this way the implementation of our own software in C-Objetive for NextStep operating system in the Microcomputer Department at the Autonomous University of Puebla in 1996. First author also acknowledges the support of EPSRC (grant EP/D066174/1) and the previous support of CONACyT with register number 139509.


\appendix

\section{Finite subset of regular expressions gliders-based}

We present the complete subset $Ph_1$ of regular expressions determining a particular phase (periodic sequence), for each glider up to now known in Rule 110.\footnote{The subset of regular expressions is also available in a text file ``listPhasesR110.txt'' from \url{http://uncomp.uwe.ac.uk/genaro/Rule110.html}}

\small

\subsection{ether}

$e$(f$_1$\_1) = 11111000100110

\subsection{$A$ glider}

$A$(f$_1$\_1) = 111110 \\
$A$(f$_2$\_1) = 11111000111000100110 \\
$A$(f$_3$\_1) = 11111000100110100110 \\
$A$(f$_4$\_1) = $A$(f$_1$\_1)

\subsection{$B$ glider}

$B$(f$_1$\_1) = 11111010 \\
$B$(f$_2$\_1) = 11111000 \\
$B$(f$_3$\_1) = 1111100010011000100110 \\
$B$(f$_4$\_1) = 11100110

\subsection{$\bar{B}$ glider}

$\bar{B}$(A,f$_1$\_1) = 1111100010110111100110 \\
$\bar{B}$(A,f$_2$\_1) = 111110001001111111001011111000100110 \\
$\bar{B}$(A,f$_3$\_1) = 111110001001101100000101111000100110 \\
$\bar{B}$(A,f$_4$\_1) = 1111110000111100100110 \\

\noindent $\bar{B}$(B,f$_1$\_1) = 1111100001000110010110 \\
$\bar{B}$(B,f$_2$\_1) = 111110001000110011101111111000100110 \\
$\bar{B}$(B,f$_3$\_1) = 111110001001100111011011100000100110 \\
$\bar{B}$(B,f$_4$\_1) = 1110110111111010000110 \\

\noindent $\bar{B}$(C,f$_1$\_1) = 1111101111110000111000 \\
$\bar{B}$(C,f$_2$\_1) = 111110001110000100011010011000100110 \\
$\bar{B}$(C,f$_3$\_1) = 111110001001101000110011111011100110 \\
$\bar{B}$(C,f$_4$\_1) = 1111100111011000111010

\subsection{$\hat{B}$ glider}

$\hat{B}$(A,f$_1$\_1) = 111110001011011110011001111111000100110 \\
$\hat{B}$(A,f$_2$\_1) = 111110001001111111001011101100000100110 \\ 
$\hat{B}$(A,f$_3$\_1) = 111110001001101100000101111011110000110 \\ 
$\hat{B}$(A,f$_4$\_1) = 1111110000111100111001000 \\

\noindent $\hat{B}$(B,f$_1$\_1) = 111110000100011001011010110011000100110 \\
$\hat{B}$(B,f$_2$\_1) = 111110001000110011101111111111011100110 \\
$\hat{B}$(B,f$_3$\_1) = 111110001001100111011011100000000111010 \\
$\hat{B}$(B,f$_4$\_1) = 1110110111111010000000110 \\

\noindent $\hat{B}$(C,f$_1$\_1)  = 111110111111000011100000011111000100110 \\
$\hat{B}$(C,f$_2$\_1) = 111110001110000100011010000011000100110 \\ 
$\hat{B}$(C,f$_3$\_1) = 111110001001101000110011111000011100110 \\
$\hat{B}$(C,f$_4$\_1) = 1111100111011000100011010

\subsection{$C_1$ glider}

$C_1$(A,f$_1$\_1) = 111110000 \\
$C_1$(A,f$_2$\_1) = 11111000100011000100110 \\
$C_1$(A,f$_3$\_1) = 11111000100110011100110 \\
$C_1$(A,f$_4$\_1) = 111011010 \\

\noindent $C_1$(B,f$_1$\_1) = 11111011111111000100110 \\
$C_1$(B,f$_2$\_1) = 11111000111000000100110 \\
$C_1$(B,f$_3$\_1) = 11111000100110100000110 \\
$C_1$(B,f$_4$\_1) = $C_1$(B,f$_1$\_1)

\subsection{$C_2$ glider}

$C_2$(A,f$_1$\_1) = 11111000000100110 \\
$C_2$(A,f$_2$\_1) = 11111000100000110 \\
$C_2$(A,f$_3$\_1) = 11111000100110000 \\
$C_2$(A,f$_4$\_1) = 11100011000100110 \\

\noindent $C_2$(B,f$_1$\_1) = 11111010011100110 \\ 
$C_2$(B,f$_2$\_1) = 11111000111011010 \\
$C_2$(B,f$_3$\_1) = 1111100010011011111111000100110 \\
$C_2$(B,f$_4$\_1) = $C_2$(B,f$_1$\_1)

\subsection{$C_3$ glider}

$C_3$(A,f$_1$\_1) = 11111011010 \\
$C_3$(A,f$_2$\_1) = 1111100011111111000100110 \\
$C_3$(A,f$_3$\_1) = 1111100010011000000100110 \\
$C_3$(A,f$_4$\_1) = 11100000110 \\

\noindent $C_3$(B,f$_1$\_1) = 11111010000 \\
$C_3$(B,f$_2$\_1) = 1111100011100011000100110 \\
$C_3$(B,f$_3$\_1) = 1111100010011010011100110 \\
$C_3$(B,f$_4$\_1) = $C_3$(B,f$_1$\_1)

\subsection{$D_1$ glider}

$D_1$(A,f$_1$\_1) = 11111000010 \\
$D_1$(A,f$_2$\_1) = 1111100010001111000100110 \\
$D_1$(A,f$_3$\_1) = 1111100010011001100100110 \\
$D_1$(A,f$_4$\_1) = 11101110110 \\

\noindent $D_1$(B,f$_1$\_1) = 1111101110111111000100110 \\
$D_1$(B,f$_2$\_1) = 1111100011101110000100110 \\
$D_1$(B,f$_3$\_1) = 1111100010011011101000110 \\
$D_1$(B,f$_4$\_1) = $D_1$(C,f$_1$\_1) \\

\noindent $D_1$(C,f$_1$\_1) = 11111011100 \\
$D_1$(C,f$_2$\_1) = 1111100011101011000100110 \\
$D_1$(C,f$_3$\_1) = 1111100010011011111100110 \\
$D_1$(C,f$_4$\_1) = $D_1$(A,f$_1$\_1)

\subsection{$D_2$ glider}

$D_2$(A,f$_1$\_1) = 1111101011000100110 \\
$D_2$(A,f$_2$\_1) = 1111100011111100110 \\
$D_2$(A,f$_3$\_1) = 1111100010011000010 \\
$D_2$(A,f$_4$\_1) = 1110001111000100110 \\

\noindent $D_2$(B,f$_1$\_1) = 1111101001100100110 \\
$D_2$(B,f$_2$\_1) = 1111100011101110110 \\
$D_2$(B,f$_3$\_1) = 111110001001101110111111000100110 \\
$D_2$(B,f$_4$\_1) = $D_2$(C,f$_1$\_1) \\

\noindent $D_2$(C,f$_1$\_1) = 1111101110000100110 \\
$D_2$(C,f$_2$\_1) = 1111100011101000110 \\
$D_2$(C,f$_3$\_1) = 1111100010011011100 \\
$D_2$(C,f$_4$\_1) = $D_2$(A,f$_1$\_1)

\subsection{$E$ glider}

$E$(A,f$_1$\_1) = 1111100000000100110 \\
$E$(A,f$_2$\_1) = 1111100010000000110 \\
$E$(A,f$_3$\_1) = 1111100010011000000 \\
$E$(A,f$_4$\_1) = 1110000011000100110 \\

\noindent $E$(B,f$_1$\_1) = 1111101000011100110 \\
$E$(B,f$_2$\_1) = 1111100011100011010 \\
$E$(B,f$_3$\_1) = 111110001001101001111111000100110 \\
$E$(B,f$_4$\_1) = $E$(C,f$_1$\_1) \\

\noindent $E$(C,f$_1$\_1) = 1111101100000100110 \\
$E$(C,f$_2$\_1) = 1111100011110000110 \\
$E$(C,f$_3$\_1) = 1111100010011001000 \\
$E$(C,f$_4$\_1) = 1110110011000100110 \\

\noindent $E$(D,f$_1$\_1) = 1111101111011100110 \\
$E$(D,f$_2$\_1) = 1111100011100111010 \\
$E$(D,f$_3$\_1) = 1111100010011010110 \\
$E$(D,f$_4$\_1) = 1111111111000100110

\subsection{$\bar{E}$ glider}

$\bar{E}$(A,f$_1$\_1) = 111110000100011111010 \\
$\bar{E}$(A,f$_2$\_1) = 111110001000110011000 \\
$\bar{E}$(A,f$_3$\_1) = 11111000100110011101110011000100110 \\
$\bar{E}$(A,f$_4$\_1) = 111011011101011100110 \\

\noindent $\bar{E}$(B,f$_1$\_1) = 111110111111011111010 \\
$\bar{E}$(B,f$_2$\_1) = 111110001110000111000 \\
$\bar{E}$(B,f$_3$\_1) = 11111000100110100011010011000100110 \\
$\bar{E}$(B,f$_4$\_1) = 111110011111011100110 \\

\noindent $\bar{E}$(C,f$_1$\_1) = 111110001011000111010 \\
$\bar{E}$(C,f$_2$\_1) = 111110001001111100110 \\
$\bar{E}$(C,f$_3$\_1) = 11111000100110110001011111000100110 \\
$\bar{E}$(C,f$_4$\_1) = 111111001111000100110 \\

\noindent $\bar{E}$(D,f$_1$\_1) = 111110000101100100110 \\
$\bar{E}$(D,f$_2$\_1) = 111110001000111110110 \\
$\bar{E}$(D,f$_3$\_1) = 11111000100110011000111111000100110 \\
$\bar{E}$(D,f$_4$\_1) = 111011100110000100110 \\

\noindent $\bar{E}$(E,f$_1$\_1) = 111110111010111000110 \\
$\bar{E}$(E,f$_2$\_1) = 111110001110111110100 \\
$\bar{E}$(E,f$_3$\_1) = 11111000100110111000111011000100110 \\
$\bar{E}$(E,f$_4$\_1) = $\bar{E}$(F,f$_1$\_1) \\

\noindent $\bar{E}$(F,f$_1$\_1) = 111110100110111100110 \\
$\bar{E}$(F,f$_2$\_1) = 111110001110111110010 \\
$\bar{E}$(F,f$_3$\_1) = 11111000100110111000101111000100110 \\
$\bar{E}$(F,f$_4$\_1) = $\bar{E}$(G,f$_1$\_1) \\

\noindent $\bar{E}$(G,f$_1$\_1) = 111110100111100100110 \\
$\bar{E}$(G,f$_2$\_1) = 111110001110110010110 \\
$\bar{E}$(G,f$_3$\_1) = 11111000100110111101111111000100110 \\
$\bar{E}$(G,f$_4$\_1) = 111110011100000100110 \\

\noindent $\bar{E}$(H,f$_1$\_1) = 111110001011010000110 \\
$\bar{E}$(H,f$_2$\_1) = 111110001001111111000 \\
$\bar{E}$(H,f$_3$\_1) = 11111000100110110000010011000100110 \\
$\bar{E}$(H,f$_4$\_1) = 111111000011011100110

\subsection{$F$ glider}

$F$(A,f$_1$\_1) = 111110001011010 \\
$F$(A,f$_2$\_1) = 11111000100111111111000100110 \\
$F$(A,f$_3$\_1) = 11111000100110110000000100110 \\
$F$(A,f$_4$\_1) = 111111000000110 \\

\noindent $F$(B,f$_1$\_1) = 111110000100000 \\
$F$(B,f$_2$\_1) = 11111000100011000011000100110 \\
$F$(B,f$_3$\_1) = 11111000100110011100011100110 \\
$F$(B,f$_4$\_1) = 111011010011010 \\

\noindent $F$(C,f$_1$\_1) = 11111011111101111111000100110 \\
$F$(C,f$_2$\_1) = 11111000111000011100000100110 \\
$F$(C,f$_3$\_1) = 11111000100110100011010000110 \\
$F$(C,f$_4$\_1) = 111110011111000 \\

\noindent $F$(D,f$_1$\_1) = 11111000101100010011000100110 \\
$F$(D,f$_2$\_1) = 11111000100111110011011100110 \\
$F$(D,f$_3$\_1) = 11111000100110110001011111010 \\
$F$(D,f$_4$\_1) = 111111001111000 \\

\noindent $F$(E,f$_1$\_1) = 11111000010110010011000100110 \\
$F$(E,f$_2$\_1) = 11111000100011111011011100110 \\
$F$(E,f$_3$\_1) = 11111000100110011000111111010 \\
$F$(E,f$_4$\_1) = 111011100110000 \\

\noindent $F$(F,f$_1$\_1) = 11111011101011100011000100110 \\
$F$(F,f$_2$\_1) = 11111000111011111010011100110 \\
$F$(F,f$_3$\_1) = 11111000100110111000111011010 \\
$F$(F,f$_4$\_1) = $F$(G,f$_1$\_1) \\

\noindent $F$(G,f$_1$\_1) = 11111010011011111111000100110 \\
$F$(G,f$_2$\_1) = 11111000111011111000000100110 \\
$F$(G,f$_3$\_1) = 11111000100110111000100000110 \\
$F$(G,f$_4$\_1) = $F$(H,f$_1$\_1) \\

\noindent $F$(H,f$_1$\_1) = 111110100110000 \\
$F$(H,f$_2$\_1) = 11111000111011100011000100110 \\
$F$(H,f$_3$\_1) = 11111000100110111010011100110 \\
$F$(H,f$_4$\_1) = $F$(A2,f$_1$\_1) \\

\noindent $F$(A2,f$_1$\_1) = 111110111011010 \\
$F$(A2,f$_2$\_1) = 11111000111011111111000100110 \\
$F$(A2,f$_3$\_1) = 11111000100110111000000100110 \\
$F$(A2,f$_4$\_1) = $F$(B2,f$_1$\_1) \\

\noindent $F$(B2,f$_1$\_1) = 111110100000110 \\
$F$(B2,f$_2$\_1) = 111110001110000 \\
$F$(B2,f$_3$\_1) = 11111000100110100011000100110 \\
$F$(B2,f$_4$\_1) = 111110011100110

\subsection{$G$ glider}

$G$(A,f$_1$\_1) = 111110100111110011100110 \\
$G$(A,f$_2$\_1) = 111110001110110001011010 \\
$G$(A,f$_3$\_1) = 11111000100110111100111111111000100110 \\
$G$(A,f$_4$\_1) = 111110010110000000100110 \\

\noindent $G$(B,f$_1$\_1) = 111110001011111000000110 \\
$G$(B,f$_2$\_1) = 111110001001111000100000 \\
$G$(B,f$_3$\_1) = 11111000100110110010011000011000100110 \\
$G$(B,f$_4$\_1) = 111111011011100011100110 \\

\noindent $G$(C,f$_1$\_1) = 111110000111111010011010 \\
$G$(C,f$_2$\_1) = 11111000100011000011101111111000100110 \\
$G$(C,f$_3$\_1) = 11111000100110011100011011100000100110 \\
$G$(C,f$_4$\_1) = 111011010011111010000110 \\

\noindent $G$(D,f$_1$\_1) = 111110111111011000111000 \\
$G$(D,f$_2$\_1) = 11111000111000011110011010011000100110 \\
$G$(D,f$_3$\_1) = 11111000100110100011001011111011100110 \\
$G$(D,f$_4$\_1) = 111110011101111000111010 \\

\noindent $G$(E,f$_1$\_1) = 111110001011011100100110 \\
$G$(E,f$_2$\_1) = 11111000100111111101011011111000100110 \\
$G$(E,f$_3$\_1) = 11111000100110110000011111111000100110 \\
$G$(E,f$_4$\_1) = 111111000011000000100110 \\

\noindent $G$(F,f$_1$\_1) = 111110000100011100000110 \\
$G$(F,f$_2$\_1) = 11111000100011001101000011111000100110 \\
$G$(F,f$_3$\_1) = 11111000100110011101111100011000100110 \\
$G$(F,f$_4$\_1) = 111011011100010011100110 \\

\noindent $G$(G,f$_1$\_1) = 111110111111010011011010 \\
$G$(G,f$_2$\_1) = 11111000111000011101111111111000100110 \\
$G$(G,f$_3$\_1) = 11111000100110100011011100000000100110 \\
$G$(G,f$_4$\_1) = 111110011111010000000110 \\

\noindent $G$(H,f$_1$\_1) = 111110001011000111000000 \\
$G$(H,f$_2$\_1) = 11111000100111110011010000011000100110 \\
$G$(H,f$_3$\_1) = 11111000100110110001011111000011100110 \\
$G$(H,f$_4$\_1) = 111111001111000100011010 \\

\noindent $G$(A2,f$_1$\_1) = 11111000010110010011001111111000100110 \\
$G$(A2,f$_2$\_1) = 11111000100011111011011101100000100110 \\
$G$(A2,f$_3$\_1) = 11111000100110011000111111011110000110 \\
$G$(A2,f$_4$\_1) = 111011100110000111001000 \\

\noindent $G$(B2,f$_1$\_1) = 11111011101011100011010110011000100110 \\
$G$(B2,f$_2$\_1) = 11111000111011111010011111111011100110 \\
$G$(B2,f$_3$\_1) = 11111000100110111000111011000000111010 \\
$G$(B2,f$_4$\_1) = $G$(C2,f$_1$\_1) \\

\noindent $G$(C2,f$_1$\_1) = 111110100110111100000110 \\
$G$(C2,f$_2$\_1) = 11111000111011111001000011111000100110 \\
$G$(C2,f$_3$\_1) = 11111000100110111000101100011000100110 \\
$G$(C2,f$_4$\_1) = $G$(A,f$_1$\_1)

\subsection{$H$ glider}

$H$(A,f$_1$\_1) = 11111000101100000000111110001001101001111111000100110 \\
$H$(A,f$_2$\_1) = 11111000100111110000000110001001101111101100000100110 \\
$H$(A,f$_3$\_1) = 11111000100110110001000000111001101111100011110000110 \\
$H$(A,f$_4$\_1) = 111111001100000110101111100010011001000 \\

\noindent $H$(B,f$_1$\_1) = 11111000010111000011111110001001101110110011000100110 \\
$H$(B,f$_2$\_1) = 11111000100011110100011000001001101111101111011100110 \\
$H$(B,f$_3$\_1) = 11111000100110011001110011100001101111100011100111010 \\
$H$(B,f$_4$\_1) = 111011101101011010001111100010011010110 \\

\noindent $H$(C,f$_1$\_1) = 11111011101111111111100110001001101111111111000100110 \\
$H$(C,f$_2$\_1) = 11111000111011100000000010111001101111100000000100110 \\
$H$(C,f$_3$\_1) = 11111000100110111010000000011110101111100010000000110 \\
$H$(C,f$_4$\_1) = $H$(D,f$_1$\_1) \\

\noindent $H$(D,f$_1$\_1) = 111110111000000011001111100010011000000 \\
$H$(D,f$_2$\_1) = 11111000111010000001110110001001101110000011000100110 \\
$H$(D,f$_3$\_1) = 11111000100110111000001101111001101111101000011100110 \\
$H$(D,f$_4$\_1) = $H$(E,f$_1$\_1) \\

\noindent $H$(E,f$_1$\_1) = 111110100001111100101111100011100011010 \\
$H$(E,f$_2$\_1) = 11111000111000110001011110001001101001111111000100110 \\
$H$(E,f$_3$\_1) = 11111000100110100111001111001001101111101100000100110 \\
$H$(E,f$_4$\_1) = $H$(F,f$_1$\_1) \\

\noindent $H$(F,f$_1$\_1) = 111110110101100101101111100011110000110 \\
$H$(F,f$_2$\_1) = 111110001111111110111111100010011001000 \\
$H$(F,f$_3$\_1) = 11111000100110000000111000001001101110110011000100110 \\
$H$(F,f$_4$\_1) = 111000000110100001101111101111011100110 \\

\noindent $H$(G,f$_1$\_1) = 111110100000111110001111100011100111010 \\
$H$(G,f$_2$\_1) = 111110001110000110001001100010011010110 \\
$H$(G,f$_3$\_1) = 11111000100110100011100110111001101111111111000100110 \\
$H$(G,f$_4$\_1) = 111110011010111110101111100000000100110 \\

\noindent $H$(H,f$_1$\_1) = 111110001011111110001111100010000000110 \\
$H$(H,f$_2$\_1) = 111110001001111000001001100010011000000 \\
$H$(H,f$_3$\_1) = 11111000100110110010000110111001101110000011000100110 \\
$H$(H,f$_4$\_1) = 111111011000111110101111101000011100110 \\

\noindent $H$(A2,f$_1$\_1) = 111110000111100110001111100011100011010 \\
$H$(A2,f$_2$\_1) = 11111000100011001011100110001001101001111111000100110 \\
$H$(A2,f$_3$\_1) = 11111000100110011101111010111001101111101100000100110 \\
$H$(A2,f$_4$\_1) = 111011011100111110101111100011110000110 \\

\noindent $H$(B2,f$_1$\_1) = 111110111111010110001111100010011001000 \\
$H$(B2,f$_2$\_1) = 11111000111000011111100110001001101110110011000100110 \\
$H$(B2,f$_3$\_1) = 11111000100110100011000010111001101111101111011100110 \\
$H$(B2,f$_4$\_1) = 111110011100011110101111100011100111010 \\

\noindent $H$(C2,f$_1$\_1) = 111110001011010011001111100010011010110 \\
$H$(C2,f$_2$\_1) = 11111000100111111101110110001001101111111111000100110 \\
$H$(C2,f$_3$\_1) = 11111000100110110000011101111001101111100000000100110 \\
$H$(C2,f$_4$\_1) = 111111000011011100101111100010000000110 \\

\noindent $H$(D2,f$_1$\_1) = 111110000100011111010111100010011000000 \\
$H$(D2,f$_2$\_1) = 11111000100011001100011111001001101110000011000100110 \\
$H$(D2,f$_3$\_1) = 11111000100110011101110011000101101111101000011100110 \\
$H$(D2,f$_4$\_1) = 111011011101011100111111100011100011010 \\

\noindent $H$(E2,f$_1$\_1) = 11111011111101111101011000001001101001111111000100110 \\
$H$(E2,f$_2$\_1) = 11111000111000011100011111100001101111101100000100110 \\
$H$(E2,f$_3$\_1) = 11111000100110100011010011000010001111100011110000110 \\
$H$(E2,f$_4$\_1) = 111110011111011100011001100010011001000 \\

\noindent $H$(F2,f$_1$\_1) = 11111000101100011101001110111001101110110011000100110 \\
$H$(F2,f$_2$\_1) = 11111000100111110011011101101110101111101111011100110 \\
$H$(F2,f$_3$\_1) = 11111000100110110001011111011111101111100011100111010 \\
$H$(F2,f$_4$\_1) = 111111001111000111000011100010011010110 \\

\noindent $H$(G2,f$_1$\_1) = 11111000010110010011010001101001101111111111000100110 \\
$H$(G2,f$_2$\_1) = 11111000100011111011011111001111101111100000000100110 \\
$H$(G2,f$_3$\_1) = 11111000100110011000111111000101100011100010000000110 \\
$H$(G2,f$_4$\_1) = 111011100110000100111110011010011000000 \\

\noindent $H$(H2,f$_1$\_1) = 11111011101011100011011000101111101110000011000100110 \\
$H$(H2,f$_2$\_1) = 11111000111011111010011111100111100011101000011100110 \\
$H$(H2,f$_3$\_1) = 11111000100110111000111011000010110010011011100011010 \\
$H$(H2,f$_4$\_1) = $H$(A3,f$_1$\_1) \\

\noindent $H$(A3,f$_1$\_1) = 11111010011011110001111101101111101001111111000100110 \\
$H$(A3,f$_2$\_1) = 11111000111011111001001100011111100011101100000100110 \\
$H$(A3,f$_3$\_1) = 11111000100110111000101101110011000010011011110000110 \\
$H$(A3,f$_4$\_1) = $H$(B3,f$_1$\_1) \\

\noindent $H$(B3,f$_1$\_1) = 111110100111111101011100011011111001000 \\
$H$(B3,f$_2$\_1) = 11111000111011000001111101001111100010110011000100110 \\
$H$(B3,f$_3$\_1) = 11111000100110111100001100011101100010011111011100110 \\
$H$(B3,f$_4$\_1) = 111110010001110011011110011011000111010 \\

\noindent $H$(C3,f$_1$\_1) = 111110001011001101011111001011111100110 \\
$H$(C3,f$_2$\_1) = 11111000100111110111111100010111100001011111000100110 \\
$H$(C3,f$_3$\_1) = 11111000100110110001110000010011110010001111000100110 \\
$H$(C3,f$_4$\_1) = 111111001101000011011001011001100100110 \\

\noindent $H$(D3,f$_1$\_1) = 111110000101111100011111101111101110110 \\
$H$(D3,f$_2$\_1) = 11111000100011110001001100001110001110111111000100110 \\
$H$(D3,f$_3$\_1) = 11111000100110011001001101110001101001101110000100110 \\
$H$(D3,f$_4$\_1) = 111011101101111101001111101111101000110 \\

\noindent $H$(E3,f$_1$\_1) = 111110111011111100011101100011100011100 \\
$H$(E3,f$_2$\_1) = 11111000111011100001001101111001101001101011000100110 \\
$H$(E3,f$_3$\_1) = 11111000100110111010001101111100101111101111111100110 \\
$H$(E3,f$_4$\_1) = $H$(F3,f$_1$\_1) \\

\noindent $H$(F3,f$_1$\_1) = 111110111001111100010111100011100000010 \\
$H$(F3,f$_2$\_1) = 11111000111010110001001111001001101000001111000100110 \\
$H$(F3,f$_3$\_1) = 11111000100110111111001101100101101111100001100100110 \\
$H$(F3,f$_4$\_1) = $H$(G3,f$_1$\_1) \\

\noindent $H$(G3,f$_1$\_1) = 111110000101111110111111100010001110110 \\
$H$(G3,f$_2$\_1) = 11111000100011110000111000001001100110111111000100110 \\
$H$(G3,f$_3$\_1) = 11111000100110011001000110100001101110111110000100110 \\
$H$(G3,f$_4$\_1) = 111011101100111110001111101110001000110 \\

\noindent $H$(H3,f$_1$\_1) = 111110111011110110001001100011101001100 \\
$H$(H3,f$_2$\_1) = 11111000111011100111100110111001101110111011000100110 \\
$H$(H3,f$_3$\_1) = 11111000100110111010110010111110101111101110111100110 \\
$H$(H3,f$_4$\_1) = $H$(A4,f$_1$\_1) \\

\noindent $H$(A4,f$_1$\_1) = 111110111111011110001111100011101110010 \\
$H$(A4,f$_2$\_1) = 11111000111000011100100110001001101110101111000100110 \\
$H$(A4,f$_3$\_1) = 11111000100110100011010110111001101111101111100100110 \\
$H$(A4,f$_4$\_1) = 111110011111111110101111100011100010110

\subsection{Glider gun}

gun(A,f$_1$\_1) = 11111010110011101001100101111100000100110 \\
gun(A,f$_2$\_1) = 11111000111111011011101110111100010000110 \\
gun(A,f$_3$\_1) = 11111000100110000111111011101110010011000 \\
gun(A,f$_4$\_1) = 11100011000011101110101101110011000100110 \\

\noindent gun(B,f$_1$\_1) = 11111010011100011011101111111101011100110 \\
gun(B,f$_2$\_1) = 11111000111011010011111011100000011111010 \\
gun(B,f$_3$\_1) = 11111000100110111111011000111010000011000 \\
gun(B,f$_4$\_1) = gun(C,f$_1$\_1) \\

\noindent gun(C,f$_1$\_1) = 11111000011110011011100001110011000100110 \\
gun(C,f$_2$\_1) = 11111000100011001011111010001101011100110 \\
gun(C,f$_3$\_1) = 11111000100110011101111000111001111111010 \\
gun(C,f$_4$\_1) = 111011011100100110101100000 \\

\noindent gun(D,f$_1$\_1) = 11111011111101011011111111000011000100110 \\
gun(D,f$_2$\_1) = 11111000111000011111111000000100011100110 \\
gun(D,f$_3$\_1) = 11111000100110100011000000100000110011010 \\
gun(D,f$_4$\_1) = 11111001110000011000011101111111000100110 \\

\noindent gun(E,f$_1$\_1) = 11111000101101000011100011011100000100110 \\
gun(E,f$_2$\_1) = 11111000100111111100011010011111010000110 \\
gun(E,f$_3$\_1) = 11111000100110110000010011111011000111000 \\
gun(E,f$_4$\_1) = 11111100001101100011110011010011000100110 \\

\noindent gun(F,f$_1$\_1) = 11111000010001111110011001011111011100110 \\
gun(F,f$_2$\_1) = 11111000100011001100001011101111000111010 \\
gun(F,f$_3$\_1) = 11111000100110011101110001111011100100110 \\
gun(F,f$_4$\_1) = 11101101110100110011101011011111000100110 \\

\noindent gun(G,f$_1$\_1) = 11111011111101110111011011111111000100110 \\
gun(G,f$_2$\_1) = 11111000111000011101110111111000000100110 \\
gun(G,f$_3$\_1) = 11111000100110100011011101110000100000110 \\
gun(G,f$_4$\_1) = 11111001111101110100011000011111000100110 \\

\noindent gun(H,f$_1$\_1) = 11111000101100011101110011100011000100110 \\
gun(H,f$_2$\_1) = 11111000100111110011011101011010011100110 \\
gun(H,f$_3$\_1) = 11111000100110110001011111011111111011010 \\
gun(H,f$_4$\_1) = 11111100111100011100000011111111000100110 \\

\noindent gun(A2,f$_1$\_1) = 11111000010110010011010000011000000100110 \\
gun(A2,f$_2$\_1) = 11111000100011111011011111000011100000110 \\
gun(A2,f$_3$\_1) = 11111000100110011000111111000100011010000 \\
gun(A2,f$_4$\_1) = 11101110011000010011001111100011000100110 \\

\noindent gun(B2,f$_1$\_1) = 11111011101011100011011101100010011100110 \\
gun(B2,f$_2$\_1) = 11111000111011111010011111011110011011010 \\
gun(B2,f$_3$\_1) = 1111100010011011100011101100011100101111111111000100110 \\
gun(B2,f$_4$\_1) = gun(C2,f$_1$\_1) \\

\noindent gun(C2,f$_1$\_1) = 11111010011011110011010111100000000100110 \\
gun(C2,f$_2$\_1) = 11111000111011111001011111110010000000110 \\
gun(C2,f$_3$\_1) = 11111000100110111000101111000001011000000 \\
gun(C2,f$_4$\_1) = gun(D2,f$_1$\_1) \\

\noindent gun(D2,f$_1$\_1) = 11111010011110010000111110000011000100110 \\
gun(D2,f$_2$\_1) = 11111000111011001011000110001000011100110 \\
gun(D2,f$_3$\_1) = 11111000100110111101111100111001100011010 \\
gun(D2,f$_4$\_1) = 11111001110001011010111001111111000100110 \\

\noindent gun(E2,f$_1$\_1) = 11111000101101001111111110101100000100110 \\
gun(E2,f$_2$\_1) = 11111000100111111101100000001111110000110 \\
gun(E2,f$_3$\_1) = 11111000100110110000011110000001100001000 \\
gun(E2,f$_4$\_1) = 11111100001100100000111000110011000100110 \\

\noindent gun(F2,f$_1$\_1) = 11111000010001110110000110100111011100110 \\
gun(F2,f$_2$\_1) = 11111000100011001101111000111110110111010 \\
gun(F2,f$_3$\_1) = 11111000100110011101111100100110001111110 \\
gun(F2,f$_4$\_1) = 11101101110001011011100110000111000100110 \\

\noindent gun(G2,f$_1$\_1) = 11111011111101001111111010111000110100110 \\
gun(G2,f$_2$\_1) = 11111000111000011101100000111110100111110 \\
gun(G2,f$_3$\_1) = 1111100010011010001101111000011000111011000111000100110 \\
gun(G2,f$_4$\_1) = 11111001111100100011100110111100110100110 \\

\noindent gun(H2,f$_1$\_1) = 11111000101100010110011010111110010 \\
gun(H2,f$_2$\_1) = 1111100010011111001111101111111000101111000100110 \\
gun(H2,f$_3$\_1) = 1111100010011011000101100011100000100111100100110 \\
gun(H2,f$_4$\_1) = 11111100111110011010000110110010110 \\

\noindent gun(A3,f$_1$\_1) = 1111100001011000101111100011111101111111000100110 \\
gun(A3,f$_2$\_1) = 1111100010001111100111100010011000011100000100110 \\
gun(A3,f$_3$\_1) = 1111100010011001100010110010011011100011010000110 \\
gun(A3,f$_4$\_1) = 11101110011111011011111010011111000 \\

\noindent gun(B3,f$_1$\_1) = 1111101110101100011111100011101100010011000100110 \\
gun(B3,f$_2$\_1) = 1111100011101111110011000010011011110011011100110 \\
gun(B3,f$_3$\_1) = 1111100010011011100001011100011011111001011111010 \\
gun(B3,f$_4$\_1) = gun(C3,f$_1$\_1) \\

\noindent gun(C3,f$_1$\_1) = 11111010001111010011111000101111000 \\
gun(C3,f$_2$\_1) = 1111100011100110011101100010011110010011000100110 \\
gun(C3,f$_3$\_1) = 1111100010011010111011011110011011001011011100110 \\
gun(C3,f$_4$\_1) = 11111110111111001011111101111111010 \\

\noindent gun(D3,f$_1$\_1) = 11111000001110000101111000011100000 \\
gun(D3,f$_2$\_1) = 1111100010000110100011110010001101000011000100110 \\
gun(D3,f$_3$\_1) = 1111100010011000111110011001011001111100011100110 \\
gun(D3,f$_4$\_1) = 11100110001011101111101100010011010 \\

\noindent gun(E3,f$_1$\_1) = 1111101011100111101110001111001101111111000100110

\end{document}